\documentclass[1p]{elsarticle}

\makeatletter
\def\ps@pprintTitle{%
	\let\@oddhead\@empty
	\let\@evenhead\@empty
	\def\@oddfoot{}%
	\let\@evenfoot\@oddfoot}
\makeatother

\usepackage[utf8]{inputenc}
\usepackage{lineno,hyperref}
\usepackage{amssymb}
\usepackage{mathtools}
\usepackage{mathpazo}
\usepackage{float}
\usepackage{booktabs}
\usepackage{bm}
\usepackage{enumitem}
\usepackage{listings}
\usepackage{caption}
\usepackage{multicol}
\usepackage{color}

\definecolor{cblue}{RGB}{68,122,187}

\newcommand{\wiser}{\textsc{Wiser}}

\usepackage{todonotes}

\usepackage{multirow}

\modulolinenumbers[5]

\let\svthefootnote\thefootnote

\begin{document}
	
	\begin{frontmatter}
		
		\title{\wiser: A Semantic Approach for Expert Finding in Academia based on Entity Linking}
		
		\author{Paolo Cifariello}
		\author{Paolo Ferragina}
		\author{Marco Ponza}
		\address{Dipartimento di Informatica, University of Pisa}

		\begin{abstract}
			We present \wiser{}, a new semantic search engine for expert finding in academia. Our system is unsupervised and it jointly combines classical language modeling techniques, based on text evidences, with the Wikipedia Knowledge Graph, via entity linking. 
			
			\wiser{} indexes each academic author through a {\em novel profiling technique} which models her expertise with a small, labeled and weighted graph drawn from Wikipedia. Nodes in this graph are the Wikipedia entities mentioned in the author's publications, whereas the weighted edges express the semantic relatedness among these entities computed via textual and graph-based relatedness functions. Every node is also labeled with a {\em relevance score} which models the pertinence of the corresponding entity to author's expertise, and is computed by means of a proper random-walk calculation over that graph; and with a {\em latent vector representation} which is learned via entity and other kinds of structural embeddings derived from Wikipedia. 
			
			At query time, experts are retrieved by combining classic document-centric appro\-ac\-h\-es, which exploit the occurrences of query terms in the author's documents, with a novel set of profile-centric scoring strategies, which compute the {\em semantic relatedness} between the author's expertise and the query topic via the above graph-based profiles.
			
			The effectiveness of our system is established over a large-scale experimental test on a standard dataset for this task. We show that \wiser{} achieves better performance than all the other competitors, thus proving the effectiveness of modeling author's profile via our ``semantic'' graph of entities. Finally, we comment on the use of \wiser{} for indexing and profiling the whole research community within the University of Pisa, and its application to technology transfer in our University.
		\end{abstract}
		
		\begin{keyword}
			expert finding \sep expert profiling \sep expertise retrieval \sep entity linking \sep information retrieval \sep wikipedia
		\end{keyword}
		
	\end{frontmatter}

	\section{Introduction}
	
	\let\thefootnote\relax\footnotetext{This work has been published at \textit{Information Systems, Elsevier (2019)} and it is available at the address \href{https://doi.org/10.1016/j.is.2018.12.003}{https://doi.org/10.1016/j.is.2018.12.003}.}\let\thefootnote\svthefootnote Searching the human expertise has  recently attracted considerable attention in the information retrieval (IR) community. This is a computationally challenging task because {\em human expertise} is hard to formalize. Expertise has been commonly referred as ``tacit knowledge'' \cite{baumard:tacit-knowledge}: namely, it is the knowledge that people carry in their minds and, therefore, it difficult to access. As a consequence, an expert finding system has one way to assess and access the expertise of a person: through artifacts of the so-called ``explicit knowledge'': namely, something that is already captured, documented, stored via e.g. documents. Applications of those systems concern with the recognition of qualified experts to supervise new researchers, assigning a paper or a project to reviewers \cite{Roberts2016}, finding relevant experts in social networks \cite{Neshati:2014:ISS:2662809.2662827} or, more important for modern academia, establishing links with industries for technology transfer initiatives. 
	
	Research on how to provide a way to actually share expertise can be traced back to at least the 1960s \cite{Brittain1975}. In more recent years, the explosion of digital information has revamped the scientific interest on this problem and led researchers to study and design software systems that, given a topic $X$, could support the {\em automatic search} for candidates with the expertise $X$. Initial approaches were mainly technical and focused on how to unify disparate and dissimilar document collections and databases into a single data warehouse that could easily be mined. They employed some heuristics or even required people to self-judge their skills against a predefined set of keywords \cite{Ackerman:2002:SEB:572524,DBLP:conf/wetice/YimamK00}. Subsequent approaches have been proposed to exploit techniques proper of document retrieval, and they have been applied to the documents written by or associated to each expert candidate as the main evidence of her expertise \cite{Balog:expertise-retrieval}. However, classical search engine return documents not people or topics \cite{oro23640}. 
	
	Today they do exist advanced systems which may be classified into two main categories: \textit{expert finding systems}, which help to find who is an expert on some topic, and \textit{expert profiling systems}, which help to find of which topic a person is an expert.  Balog et al. \cite{Balog:expertise-retrieval} summarize the general frameworks that have been used to solve these two tasks (see also Section \ref{sec:relwork}), and look at them as two sides of the same coin: an author retrieved as expert of a topic should contain that topic in her profile. However, as pointed out in \cite{DBLP:journals/corr/GyselRW16,VanGysel2016products}, known systems are yet poor in addressing three key challenges which, in turn, limit their efficiency and applicability \cite{li2014semantic,Balog:expertise-retrieval}: (1) Queries and documents use different representations so that maximum-likelihood language models are often inappropriate, and thus there is the need to make use of semantic similarities between words; (2) The acceleration of data availability calls for the further development of unsupervised methods; (3) In some approaches, a language model is constructed for every document in the collection thus requiring to match each query term against every document. 
	
	In this paper we focus on the task of {\em experts finding in the academia domain}, namely, we wish to retrieve academic authors whose expertise is defined through the publications they wrote and it is relevant for a user query. 

	In this context, the best system to date is the one recently proposed by Van Gysel et al. \cite{DBLP:journals/corr/GyselRW16}. It has a strong emphasis on unsupervised profile construction, efficient query capabilities and \textit{semantic matching} between query terms and candidate profiles. Van Gysel et al.  have shown that their unsupervised approach improves retrieval performance of vector space-based and generative-language models, mainly due to its ability to learn a \textit{profile-centric} latent representation of academic experts from their publications. Their key idea is to deploy an embedding representation of words (such as the one proposed in~\cite{mikolov2013distributed}) to map conceptually similar phrases into geometrically \textit{close} vectors (i.e. ``nyt'' is mapped into a vector close to the one of ``New York Times''). At query time, their system first maps the user query into the same latent space of experts' profiles and, then, retrieves the experts showing the highest dot-product between the embeddings of their profile and the one of the query. This way the system can efficiently address the \textit{mismatch problem} between the ``language'' of the user query and the ``language'' of authors' documents: i.e., an expert can be identified even if her documents do not contain any terms of the input query~\cite{li2014semantic}.
	
	But, despite these recent improvements, the \textit{semantic matching} implemented by Van Gysel et al. \cite{DBLP:journals/corr/GyselRW16} is yet limited to the use of {\em latent} concepts, namely ones that cannot be explicitly defined and thus cannot {\em explain} the why an expert profile matches a user query. In this paper we propose a novel approach for expert finding which is still unsupervised but, unlike \cite{DBLP:journals/corr/GyselRW16}, takes advantage of the recent IR trends in the deployment of Knowledge Graphs \cite{DBLP:journals/debu/WeikumHS16,DBLP:journals/sigir/DietzXM17} which allow modern search engines and IR tools to be more powerful in {\em semantically matching} queries to documents and allow to {\em explicitly represent concepts} occurring in those documents, as well-defined nodes in these graphs. More specifically, our approach models the academic expertise of a researcher both syntactically and semantically by orchestrating a document-centric approach, that deploys an open-source search engine (namely {\sc ElasticSearch}), and a profile-centric approach, that models in an innovative way the individual expert's knowledge not just as a list of words or a vector of latent concepts (as in \cite{DBLP:journals/corr/GyselRW16}) but as a small \textit{labeled and weighted graph} derived from Wikipedia, which is the best known and open Knowledge Graph to date. That graph will consist of labeled nodes, which are the entities mentioned in author's publications (detected via \textsc{TagMe}~\cite{Ferragina:2010:TOA:1871437.1871689}, one of the most effective entity linking systems to date), and edges weighted by means of proper entity-relatedness scores (computed via an advanced framework \cite{ponza2017two}). Moreover, every node is labeled with a {\em relevance score} which models the pertinence of the corresponding entity to author's expertise, and is computed by means of proper random-walk calculation over the author's graph; and with a {\em latent vector representation} which is learned via entity and other kinds of structural embeddings, that are derived from Wikipedia and result different from the ones proposed in \cite{DBLP:journals/corr/GyselRW16}. The use of this enriched graph allows to obtain a finer, explicit and more sophisticate modeling of author's expertise that is then used at query time to search and rank experts based on the {\em semantic relation} that exist between the words/entities occurring in the user query and the ones occurring in the author's graph. 
	
	This novel modelling and querying approach has been implemented in a system called \wiser{}, which has been experimented on the largest available dataset for benchmarking academia expert finding systems, i.e. TU dataset~\cite{deRijke:tu-collection}. This dataset consists of a total of 31,209 documents, authored by 977 researchers, and 1,266 test queries with a human-assessed ground-truth that assigns to each query a ranking of its best academic experts. \wiser{} shows statistically significant improvements over different ranking metrics and configurations. More precisely, our document-centric approach improves the profile-centric \texttt{Log-linear} model proposed by \cite{DBLP:journals/corr/GyselRW16} of +7.6\%, +7.4\% and +7\% over \textit{MAP}, \textit{MRR} and \textit{NDCG@100} scores. Whereas our profile-centric approach based on entity linking improves that \texttt{Log-linear} model of +2.4\% in \textit{MAP}, and achieves comparable results for the other metrics.   Then, we show that a proper combination of our document- and profile-centric approaches achieves a further improvement over the \texttt{Log-linear} model of +9.7\%, +12.6\% and +9.1\% in \textit{MAP}, in \textit{MRR} and in \textit{NDCG\@100}; and, furthermore, it improves the sophisticated \texttt{Ensemble} method of \cite{DBLP:journals/corr/GyselRW16}, which is currently the state-of-the-art, of +$5.4\%$, +$5.7\%$ and +$3.9\%$ on \textit{MAP}, \textit{MRR} and \textit{NDCG@100} metrics, respectively. This means that \wiser{} is designed upon the best {\em single model} and the best {\em combined models} today, thus resulting the state-of-the-art for the expert finding problem in academia.
	
	A publicly available version of \wiser{} is available at \url{http://wiser.d4science.org} for testing its functionalities about expert finding and expert profiling over the researchers of the University of Pisa. 
	
	The next Sections will review the main literature about expert finding solutions (Section \ref{sec:relwork}), in order to contextualize our problem and contributions; describe the design of \wiser{}, by detailing its constituting modules and their underlying algorithmic motivations (Section \ref{sec:wiser}); and finally present a systematic and large set of experiments conducted on \wiser{} and the state-of-the-art systems, in order to show our achievements (Section \ref{sec:experimental-results}) and identify new directions for future research (Section \ref{sec:conclusions}).

	\section{Related Work}
	\label{sec:relwork}
	
	We first discuss prior work on experts finding by describing the main challenges of this task and its differences with classic document retrieval. Then we move on to describe how our work differs from known experts finding (and profiling) approaches by commenting about its novel use of entity linking, relatedness measures and word/entity embeddings: techniques that we also describe in the next paragraphs. Finally, in the last part of this Section, we will concentrate on detailing the main differences between \wiser{} and the state-of-the-art system proposed by Van Gysel et al.~\cite{DBLP:journals/corr/GyselRW16}, because it is also the most similar to ours.
	
	\vspace{0.4cm}
	\textbf{Expert Finding (and Profiling).} Experts finding systems differ from classic search engines~\cite{chakrabarti2002mining,Manning:2008:IIR:1394399} in that they address the problem of finding the right \textit{person} (in contrast with the right document) with appropriate skills and knowledge specified via a user query. Preliminary attempts were made  in adapting classic search engines to this task with poor results \cite{Balog:expertise-retrieval}. The key issue to solve is how to represent the individual expert's knowledge~\cite{Macdonald:2006:VCA:1183614.1183671,Balog:formal-models-expert-finding,Balog:2009:LMF:1460927.1461012,DBLP:journals/corr/GyselRW16}. Among the several attempts, the ones that got most attention and success were the \textit{profile-centric} models~\cite{Balog:formal-models-expert-finding,DBLP:journals/corr/GyselRW16} and the \textit{document-centric} models~\cite{cao2005research,Macdonald:2006:VCA:1183614.1183671,Balog:expertise-retrieval}. The first ones work by creating a profile for each candidate according to the documents they are associated with, and then by ranking experts through a matching between the input query and their profiles. The second ones work by first retrieving documents which are relevant to the input query and then by ranking experts according to the relevance scores of their matching documents. The joint combination of these two approaches has shown recently to further improve the achievable performance \cite{balog2008combining,DBLP:journals/corr/GyselRW16}, as we will discuss further below. 
	
	Most of the solutions present in the literature are \textit{unsupervised} \cite{cao2005research,Macdonald:2006:VCA:1183614.1183671,Balog:formal-models-expert-finding,Balog:2009:LMF:1460927.1461012,DBLP:journals/corr/GyselRW16} since they do not need any training data for the deployment of their models. \textit{Supervised} approaches~\cite{macdonald2011learning,moreira2015learning} have been also proposed, but their application has usually been confined to data collections in which query-expert pairs are available for training~\cite{sorg2011finding,fang2010discriminative}. This is clearly a limitation that has indeed led researchers to concentrate mainly onto unsupervised approaches.

	The focus of our work is onto the design of \textit{unsupervised academia experts finding} solutions which aim at retrieving experts (i.e. academic authors) whose expertise is properly defined through the publications they wrote.  Among the most popular \textit{academic} expert finding solutions we have ArnetMiner~\cite{Tang:2008:AEM:1401890.1402008}, a system for mining academic social networks which automatically crawls and indexes research papers from the Web. Its technology relies on a probabilistic framework based on topic modeling for addressing both author ambiguity and expert ranking. Unfortunately, the implementation of the system is not publicly available and it has not been experimented on publicly available datasets. Similar comments hold true for the Scival system by Elsevier.\footnote{See \url{https://www.scival.com}.}

	Among the publicly available systems for academia expert finding, the state-of-the-art is the one recently proposed by Van Gysel et al. \cite{DBLP:journals/corr/GyselRW16}. It adapts a collection of unsupervised {\em neural-based} retrieval algorithms \cite{VanGysel2017sert}, originally deployed on product search \cite{VanGysel2016products}, to the experts finding context via a log-linear model which learns a \textit{profile-centric} latent representation of academic experts from the dataset at hand. At query time, the retrieval of experts is computed by first mapping the user query into the same latent space of experts profiles and, then, by retrieving the experts with the highest dot-product between their profile and the query. 
	
	Before discussing the differences between this approach and ours, we need to recall few technicalities regarding the main modules we will use in our solution.
	
	\vspace{0.2cm}
	\textbf{Entity Linking.} All expert finding approaches mentioned above (as well as typical IR solutions to indexing, clustering and classification) are commonly based on the bag-of-words paradigm. In the last years Research went beyond this paradigm \cite{DBLP:journals/debu/WeikumHS16,DBLP:journals/sigir/DietzXM17,Ferragina:2010:TOA:1871437.1871689} with the goal of improving the search experience on unstructured or semi-structured textual data~\cite{blanco2015fast,bovi2015large,DBLP:conf/nldb/PonzaFP17,Cornolti:2016:PSJ:2872427.2883061,nguyen2017query}. The key idea is to identify sequences of terms (also called spots or mentions) in the input text and to annotate them with unambiguous entities drawn from a Knowledge Graph, such as Wikipedia \cite{DBLP:conf/emnlp/Cucerzan07,iitb}, YAGO~\cite{suchanek2007yago}, Freebase~\cite{bollacker2008freebase} or BabelNet~\cite{navigli2012babelnet}. Documents are then retrieved, classified, or clustered based on this novel representation which consists of a {\em bag of entities} and a {\em semantic relatedness} function~\cite{gabrilovich2007computing,milne-relatedness,ponza2017two} which incorporates into a floating-point number how much two entities are semantically close to each other. This novel representation has recently allowed researchers to design new algorithms that significantly boost the performance of known approaches in several IR applications, such as query understanding, documents clustering and classification, text mining, etc.~\cite{DBLP:journals/debu/WeikumHS16,DBLP:journals/sigir/DietzXM17,scaiella2012topical,ni2016semantic,Cornolti:2016:PSJ:2872427.2883061,ponza2017two}.
	
	\vspace{0.2cm}
	\textbf{Entity Embeddings.} Word embeddings~\cite{mikolov2013distributed} is a very recent Natural Language Processing (NLP) technique that aims at mapping words or phrases to low dimensional numerical vectors that are faster to manipulate and offer interesting distributional properties to compare and retrieve "similar" words or phrases \cite{mikolov2013distributed}. This latent representation has been recently extended~\cite{Ni:2016:SDR:2835776.2835801,DBLP:conf/kdd/PerozziAS14} to learn two different forms of representations of Wikipedia entities~\cite{ponza2017two}: (1) \textsc{Entity2Vec} \cite{Ni:2016:SDR:2835776.2835801} learns the latent representation of entities by working at textual-level over the content of Wikipedia pages, and (2) \textsc{DeepWalk} \cite{DBLP:conf/kdd/PerozziAS14} learns the latent representation of entities by working on the hyper-link structure of the Wikipedia graph via the execution  of random walks that start from a focus node (i.e. the entity to be embedded) and visit other nearby nodes (that provide its contextual knowledge). The former approach tends to declare {\em similar} two entities that co-occur within similar textual contexts, even if their textual mentions are different; the latter approach tends to declare {\em similar} two entities that are nearby in the Knowledge Graph. These are novel forms of semantic embeddings, which have been proved to be particularly effective in detecting entity relatedness~\cite{ponza2017two}.
	
	\vspace{0.2cm}
	To the best of our knowledge we are the first to design an experts finding system for the academia domain which is based on entity linking and embeddings techniques built upon the Wikipedia Knowledge Graph~\cite{DBLP:journals/corr/GyselRW16,Balog:expertise-retrieval}. The key feature of our system \wiser{} is a \textit{novel profile model for academic experts}, called \textit{Wikipedia Expertise Model}, that deploys those advanced techniques to build a small \textit{labeled and weighted graph} for each academic author. This graph will describe her individual "explicit" knowledge in terms of Wikipedia entities occurring in her publications and of their relatedness scores computed by means of Wikipedia-based interconnections and embeddings. This graph representation is then used at query time to efficiently search and rank academic experts based on the ``semantic" relation that exists between their graph model and the words and entities occurring in the user query.

	\section{Notation and Terminology}
	\label{sec:terminology}
	
	A dataset $(D, A)$ for the experts finding problem is a pair consisting of a set of documents $d \in D$ and a set of authors (candidate experts) $a \in A$. We  indicate with $D_a$ the set of documents written by author $a$.
	
	In our context an entity $e$ is a Wikipedia page. Entities are annotated in texts (both documents and queries) through the entity linker \textsc{TagMe}~\cite{Ferragina:2010:TOA:1871437.1871689}, which also provides a confidence score $\rho_e$ which expresses the semantic coherence between entity $e$ and its surrounding text in the input document. Since an entity $e$ can be mentioned many times in the documents of $a$, with possibly different values for $\rho_e$, we denote by $\rho_{e,a}$ the \textit{maximum} confidence score among all occurrences of $e$ in $D_a$'s documents. We use
	$E_{a}$ to denote the set of all entities annotated in the documents $D_a$ of author $a$.
	
	Given an entity $e$, we use $A_e$ to denote the set of authors who mention $e$ in one of their documents, $D_e$ to denote the subset of documents that mention $e$, and $D_{a,e}$ to denote the subset of documents written by author $a$ and which mention $e$. 
	
	A generic input query is indicated with $q$, $E_{q}$ will be used to denote the set of entities annotated in $q$ by \textsc{TagMe} and $D_{a,q}$ will be used to denote the subset of documents $D_a$ which are (syntactically or semantically) matched by the input query $q$.

	\section{\textsc{Wiser:} Our New Proposal}
	\label{sec:wiser}
	
	In this Section we describe \wiser{}, whose name stands for \textbf{Wi}kipedia Experti\textbf{se} \textbf{R}anking. It is a system for academia experts finding, built on top of three main tools:
	
	\begin{itemize}
		\item \textsc{Elasticsearch}\footnote{\url{https://www.elastic.co}}, an open-source software library for the full-text indexing of large data collections; 
		
		\item \textsc{TagMe}~\cite{Ferragina:2010:TOA:1871437.1871689}, a state-of-the-art entity linker for annotating Wikipedia pages mentioned in an input text;
		
		\item \textsc{WikipediaRelatedness}~\cite{ponza2017two}, a framework for the computation of several relatedness measures between Wikipedia entities. 
		
	\end{itemize}
	
	By properly orchestrating and enriching the results returned by the above three tools, \wiser{} offers both document-centric and profile-centric strategies for solving the experts finding problem, thus taking advantage of the positive features of both approaches. More specifically, \wiser{} first builds a document-centric model of the explicit knowledge of  academic experts via classic document indexing (by means of \textsc{Elasticsearch}) and entity annotation (by means of \textsc{TagMe}) of the authors' publications. Then, it derives a novel profile-centric model for each author that consists of a small, labeled and weighted graph drawn from Wikipedia. Nodes in this graph are the entities mentioned in the author's publications, whereas the weighted edges express the semantic relatedness among these entities, computed via \textsc{WikipediaRelatedness}. Every node is labeled with a {\em relevance score} which models the pertinence of the corresponding entity to author's expertise, and is computed by means of proper random-walk calculation over that graph; and with a {\em latent vector representation} which is learned via entity and other kinds of structural embeddings derived from Wikipedia. This graph-based model is called  \textit{Wikipedia Expertise Model} of an academic author (details in Section~\ref{sec:wiser-index-contruction}).
	
	At query time, \wiser{} uses proper data fusion techniques~\cite{Macdonald:2006:VCA:1183614.1183671} to combine several authors' ranking: the one derived from the documents' ranking provided by \textsc{Elasticsearch}, and others derived by means of properly defined "semantic matchings" between the query and the Wikipedia Expertise Model of each author. This way, it obtains a unique ranking of the academic experts that captures syntactically and semantically the searched expertise within the "explicit knowledge" of authors (details in Section~\ref{sec:wiser-query-time}). 
	
	The following sections will detail the specialties of our novel Wikipedia Expertise Model, and its construction and use in the two phases above.
	
	\subsection{Data indexing and experts modeling}
	\label{sec:wiser-index-contruction}
	
\begin{figure}[t!]
		\centering
		\includegraphics[scale=0.2]{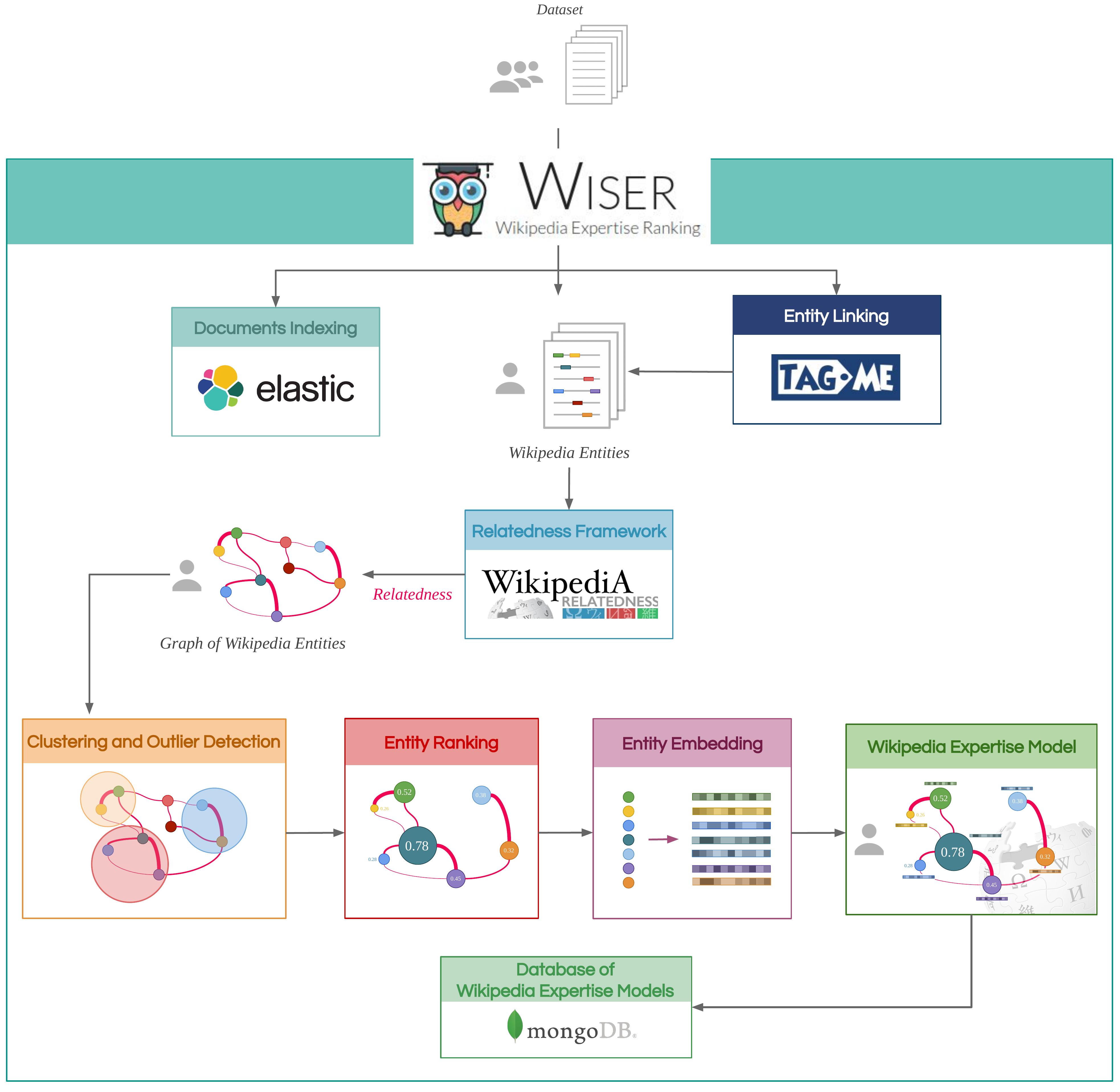}
		\caption{The construction for a given author of the Wikipedia Expertise Model in \wiser{}.}\label{fig:wiser:index}
	\end{figure}

	This is an off-line phase which consists of two main sub-phases whose goal is to construct the novel \textit{Wikipedia Expertise Model} for each academic author to be indexed. A pictorial description of this phase is provided in Figure~\ref{fig:wiser:index}.

        \begin{figure}[t]
    \centering
    \includegraphics[scale=0.23]{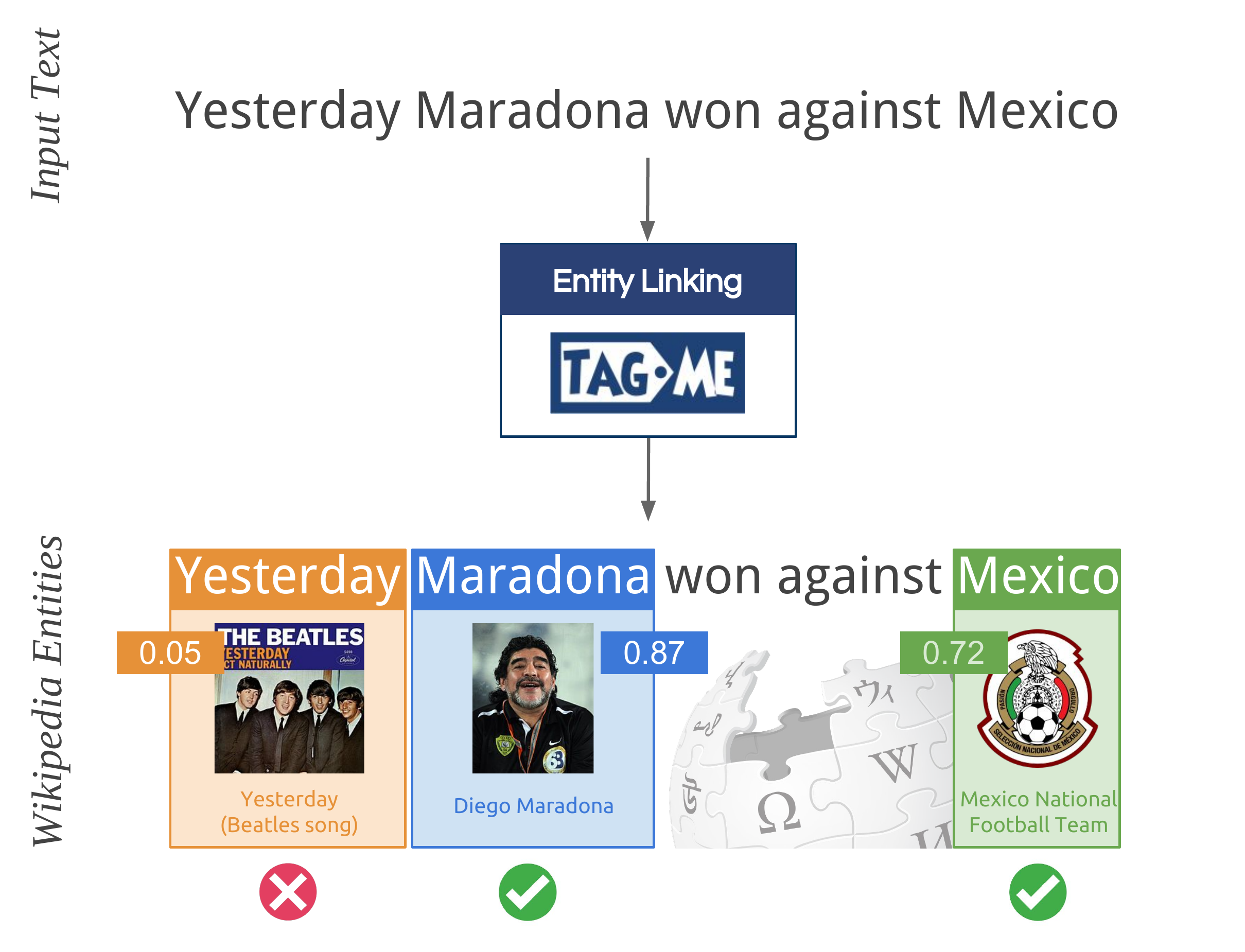}
    \caption{Example showing the benefits of filtering out entities annotated by \textsc{TagMe} with low $\rho$-scores. Both \textsf{Diego Maradona} and \textsf{Mexico National Football Team} are correctly annotated in the input text and, in fact, they receive the high $\rho$-scores of $0.87$ and $0.72$, respectively. On the other hand, \textsf{Yesterday (Beatles Song)} is wrongly annotated and, in fact, it receives a very low $\rho$-score of $0.05$.}
    \label{fig:tagme:rhoexample}
    \end{figure}

	\vspace{0.4cm}
	\noindent \textbf{Data Acquisition.} In this first sub-phase, \wiser{} indexes the authors' publications by means of \textsc{Elasticsearch} and annotates them with Wikipedia's entities by means of \textsc{TagMe}. For each input document, \textsc{Elasticsearch} stores information about its author $a$ and its textual content, whereas \textsc{TagMe} extracts the Wikipedia entities $e$ that are mentioned in the document together with their $\rho$-score that, we recall, captures the coherence between the annotated entity and the surrounding textual context in which it has been mentioned. Given that the annotated documents are scientific publications, they are well written and formatted so that \textsc{TagMe} is very effective in its task of extracting relevant Wikipedia entities. Subsequently, \wiser{} filters out the entities $e$ such that $\rho_{e,a} \leq 0.2$ (as suggested by the \textsc{TagMe}'s documentation), since those entities are usually noisy or non coherent with the topics mentioned in the annotated document (see Figure~\ref{fig:tagme:rhoexample} for an example). Eventually, all this information is stored in a MongoDB\footnote{\url{https://www.mongodb.com}} database.

	\vspace{0.4cm}
	\noindent \textbf{Wikipedia Expertise Model (abb. \textit{WEM}).} In this second sub-phase, \wiser{} creates an innovative profile of each academic author that consists of a {\em graph} whose nodes are labeled with the Wikipedia entities found in author's documents, and whose edges are weighted by deploying entity embeddings and the structure of the Wikipedia graph, by means of the \textsc{WikipediaRelatedness} framework. More precisely, the expertise of each author $a$ is modeled as a labeled and weighted graph $G_a = (V, E)$ where each node $u \in V$ is a Wikipedia entity annotated in at least one of the documents of $D_a$ by \textsc{TagMe}, and each weighted edge $(u,v) \in E$ models the relatedness between the two entities $u$ and $v$.  In our context we weight $(u,v)$ by computing the \textit{Milne\&Witten} relatedness measure between $u$'s and $v$'s entity, using the \textsc{WikipediaRelatedness} framework. This measure has shown its robustness and effectiveness in different domains~\cite{DBLP:conf/wsdm/ScaiellaFMC12,Ferragina:2010:TOA:1871437.1871689,ponza2017two}, we leave the use of more sophisticated relatedness measures, present in \textsc{WikipediaRelatedness}~\cite{ponza2017two}, to a future work.
	
	The graph $G_a$ is further refined by executing an outlier-elimination process performed via a \textit{graph clustering} algorithm that recognizes and removes from $G_a$ those entities that do not belong to any cluster and thus can be considered as off-topic for the author $a$. For this task \wiser{} deploys \textsc{HDBSCAN}~\cite{mcinnes2017accelerated}, a density-based hierarchical clustering method based on the classic \textsc{DBSCAN}~\cite{Manning:2008:IIR:1394399}. The choice in the use HDBSCAN is motivated by its efficiency and a higher clustering quality than other popular algorithms (i.e. K-Means)~\cite{mcinnes2017accelerated}. As in any clustering algorithm, input parameters of \textsc{HDBSCAN} strongly depend on the input graph and its expected output. In our experiments we observed that sometimes the entities labeled as outliers are not much off-topic (false positives), while in other cases no outliers are detected although they do exist at a human inspection (false negatives). \wiser{} deals with those issues by adopting a conservative approach: if more than 20\% of the nodes in $G_a$ are marked as outliers, we consider the output provided by HDBSCAN as not valid, and thus we keep~all~nodes in $G_a$ as valid topics for the examined author $a$.
	
	After the application of the outlier-elimination process, \wiser{} computes \textit{two attributes} for each node (hence, entity) in the graph $G_a$. The first one is the \textit{relevance} score of an entity $e$ mentioned by the author $a$. This score is computed by running the Personalized PageRank algorithm~\cite{haveliwala2002topic} over the graph $G_a$ with a proper setting of the PageRank's damping factor to $0.85$, as commonly chosen in literature~\cite{ilprints361}. Moreover, the starting and teleporting distributions over $G_a$'s nodes are defined to reflect the number of times author $a$ mentions the entity $e$ assigned to that node, and it is scaled by the $\rho$-score that evaluates how much that entity is reliable as $a$'s research topic according to {\sc TagMe}: namely, $Pr(e) = \frac{\rho_{e,a}}{C}\; \log(1 + |D_{a, e}|)$. Constant $C$ is a normalization factor that makes that formula a probability distribution over the entities labeling the nodes of $G_a$. This definition allows the more frequent  and coherent entities to get a higher chances to re-start a random walk, and thus their nodes will probably turn to get a higher steady state probability (i.e. relevance score) via the Personalized PageRank computation~\cite{haveliwala2002topic}. In this computation a significant role will be played by the weighted edges of the graph $G_a$ which explicitly model the {\em semantic relatedness} among the entities mentioned by $a$.
	
	The second attribute that is associated to each node is a \textit{vector} of floating-point numbers computed through the \textsc{DeepWalk} model for entity embeddings (see Section~\ref{sec:relwork}). This technique is inspired by the approach adopted by~\cite{VanGysel2017sert}, where the expertise of each author is modeled with an embedding vector. But, unlike~\cite{VanGysel2017sert} where vectors are learned via a bag-of-words paradigm directly from the dataset $(D, A)$, our embedding vectors are more "powerful" in that they embed the latent knowledge learned from the content and the structure of Wikipedia and, additionally, they "combine" the relevance score just described above and associated to each entity (node) in the graph $G_a$. Eventually we compute for every author $a$ one single embedding vector which is obtained by summing up the \textsc{DeepWalk}  embedding vectors relative to its top-$k$ entities and ranked according to the relevance score described above.\footnote{In the experiments of Section \ref{sec:experimental-results} we will investigate the impact of the choice of $k \in \{10, 20, 30, 50, |E_a|\}$.} This embedding vector eventually incorporates the expertise of each author into $100$ components (see Section \ref{sec:experimental-results}), thus it is fast to be managed in the subsequent query operations when we will need to compute the {\em semantic matches} between authors' topics and query topics.

	Summarizing, \wiser{} computes for every author $a$ its {\em WEM profile} which consists of the graph $G_a$ and an embedding vector of $100$ numeric components. This way the \textit{WEM} profile models the {\em explicit knowledge} of author $a$ by identifying the explicit concepts (via entities and their relations) and the latent concepts (via an embedding vector) occurring in her documents. The graph $G_a$ is crucial in many aspects because it captures the entities mentioned in $a$'s documents and their \textit{Milne\&Witten}'s relatedness score. But, also, it allows to select the top-$k$ entities that best describe $a$'s expertise, according to a relevance score derived by means of a Personalized PageRank calculation over $G_a$. The \textsc{DeepWalk} vectors of these top-$k$ entities are then summed to get the embedding vector of author $a$ that describes the best latent concepts of $a$'s expertise. 
	
	\subsection{Finding the Experts}
	\label{sec:wiser-query-time}
	
	At query time, \wiser{} operates in order to identify the expertise areas mentioned in the input query $q$ and then retrieve a set of candidate experts to which it assigns an expertise score. This score is eventually used for generating the final ranking of experts that are returned as result of query $q$.
	
\begin{figure}[t!]
		\centering
		\includegraphics[scale=0.23]{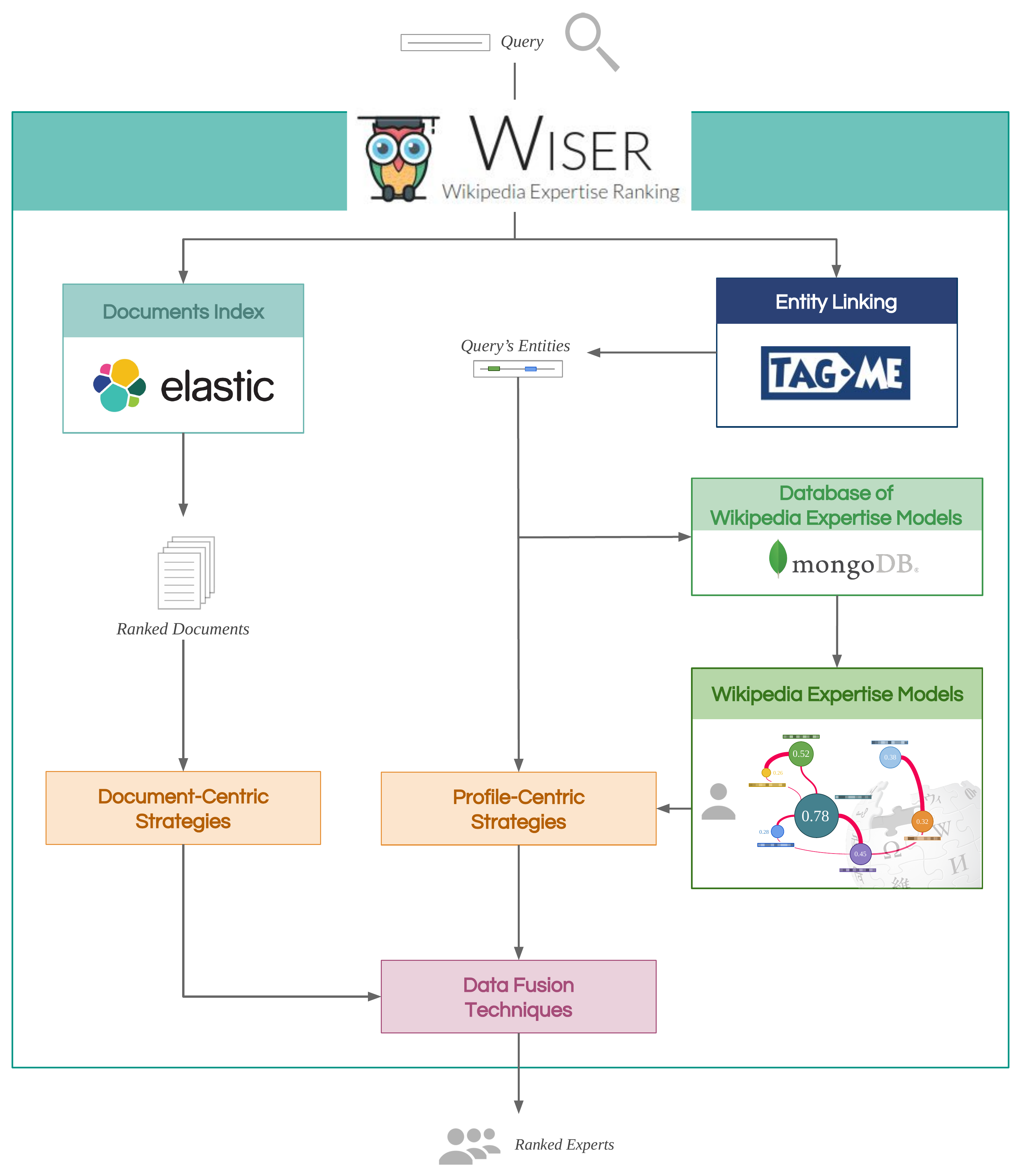}
		\caption{Experts retrieval in \wiser{} via the combination of a document-centric strategy and a profile-centric strategy through proper data fusion techniques that are described in the text.}\label{fig:wiser-query}
	\end{figure}

	Since our system relies on both \textit{document-centric} and \textit{profile-centric strategies}, we organized this Section in three main paragraphs which respectively describe each of those strategies and the method used for their combination via proper data fusion techniques. Figure \ref{fig:wiser-query} reports a graphical representation of the \textit{query} processing phase.

	\vspace{0.4cm}
	\noindent\textbf{Document-Centric Strategy.} It relies on the use of \textsc{Elasticsearch}. The query $q$ is forwarded to \textsc{Elasticsearch} in order to retrieve a ranked list of documents, i.e. a list $(d_1, s_1), \ldots, (d_n, s_n)$ where $s_i$ is the score computed for document $d_i$ given the query $q$. In our experiments we will test several ranking scores: \texttt{tf-idf}~\cite{Manning:2008:IIR:1394399}, \texttt{BM25}~\cite{robertson2009probabilistic}, and \texttt{Language Modeling} with either \texttt{Dirichlet} or \texttt{Jelinek-Mercer} smoothing ranking techniques~\cite{zhai2017study}.
	
	The ranked list of documents is then turned into a ranked list of authors $a_1, ..., a_m$ by means of several well-known techniques \cite{Macdonald:2006:VCA:1183614.1183671,Fox94combinationof} that we have adapted to our context, are described in Table~\ref{table:synquery} and tested in Section \ref{sec:experimental-results}.

	\begin{table}[ht!]
		\centering
		\caption{Document-scoring techniques used by \wiser{} within its document-centric strategy. We denote by $s_{a,j}$ the score assigned to the $j$-th document of author $a$ computed via several techniques.}
		\label{table:synquery}
		\begin{tabular}{llp{8.5cm}}
			\toprule
			\textbf{Name} & \textbf{Equation} & \textbf{Description} \\
			\hline
			\addlinespace \texttt{mean-k}  & $\frac{1}{k}\sum_{j=1}^{k} s_{a,j}$ & Average of the top-$k$ scores of $a$'s documents. \\ 
			\addlinespace \texttt{max}       &  $ \operatorname{max}( s_{a,j} )$  & Maximum of the scores of $a$'s documents.  \\
			\addlinespace \texttt{rr}       &  $ \sum_{j=1}^{|D_{a,q}|} \frac{1}{\operatorname{rank}(d_j)}$  & Reciprocal Rank~\cite{Macdonald:2006:VCA:1183614.1183671} of the ranks of $a$'s documents. $\operatorname{rank}(d_j)$ is the ranking position of document $d_j$.  \\
			\addlinespace \texttt{combnz}       &  $\frac{|D_{a,q}|}{|D_a|} \sum_{j=1}^{|D_{a,q}|} s_{a,j}$  & Documents' scores of author $a$, normalized by the number of documents associated to $a$.\\ \bottomrule
		\end{tabular}
	\end{table}

	\begin{table}[th!]
		\centering
		\caption{Author-scoring techniques based on \textit{Exact-Match} of entities and used by \wiser{} within its profile-centric strategy. The function $f$ can be linear, sigmoid or a square function. Equation \texttt{ec-iaf}, \texttt{ef-iaf} and \texttt{rec-iaf} are computed for a given author $a$ and entity $e$, whereas \texttt{max} and \texttt{mean} aggregate these scores computed for multiple entities into a single one.}
		\label{table:semquery:exactmatch}
		\begin{tabular}{llp{8.5cm}}
			\toprule
			\textbf{Name} & \textbf{Equation} & \textbf{Description} \\
			\hline
			\addlinespace \texttt{iaf} & $  \log{\frac{|A|}{|A_e|}} $ & Inverse Author Frequency, namely the \textit{smoothing factor} used for modeling the importance of entity $e$ in the dataset at hand. This score is used only when combined with other techniques (see \texttt{ec-iaf} and \texttt{ef-iaf}). \\ \hline
			\addlinespace \texttt{ec-iaf} & $ |D_{a,e}| \cdot \rho_{a,e} \cdot \texttt{iaf}(e) $ & Frequency of an entity smoothed by means of its coherence with $a$'s documents (i.e. $\rho_{a,e}$) and the {\tt iaf} scores. \\ 
			\addlinespace \texttt{ef-iaf}       & $\frac{1}{|D_a|} \cdot \texttt{ec-iaf}(a, e)$ & Scaling down \texttt{ec-iaf} by means of the "productivity" of author $a$ measured as the number $|D_a|$ of authored documents. \\
			\addlinespace \texttt{rec-iaf}       & $f(r_{a,e})  \cdot \texttt{ec-iaf}(a, e)$ & Extending \texttt{ec-iaf} equation with the relevance score $r_{a, e}$ of the entity $e$ within the graph $G_a$. $f(r_{a,e})$ is a scaling function described in the experiments. \\ \hline
			\addlinespace \texttt{max} & $\operatorname{max}( g(a, e) )$ & Maximum exact-match score computed for a given author $a \in A^q$ and for each  $e \in E_q$. $g(a, e)$ is either \texttt{ec-iaf}, \texttt{ef-iaf} or \texttt{rec-iaf}.\\
			\addlinespace \texttt{mean} & $\operatorname{mean}( g(a, e) )$ & Average exact-match score computed for a given author $a \in A^q$ and for each  $e \in E_q$. $g(a, e)$ is either \texttt{ec-iaf}, \texttt{ef-iaf} or \texttt{rec-iaf}. \\

			\bottomrule
		\end{tabular}
	\end{table}

	\newpage
	\noindent\textbf{Profile-Centric Strategy.} This is a novel set of scoring strategies that we have specifically designed to rank experts according with our new \textit{WEM} profile. Authors are scored via a computation that consists of three main steps. First, \wiser{} runs \textsc{TagMe} over the input query $q$ and annotates it with a set of pertinent Wikipedia entities, denoted by $E_q$. Second, it retrieves as candidate experts the authors $A^q$ whose \textit{WEM} profile contains at least one of the entities in $E_q$. Third, the authors in $A^q$ are ranked according to two novel {\em entity-scoring} methods, that we call \textit{Exact} and \textit{Related}, which compute authors' scores based on some properly defined \textit{exact-} or \textit{related-}scoring functions that are computed between $q$ and their \textit{WEM} profiles. These many scoring functions will be experimentally tested in Section \ref{sec:experimental-results}.
	
	\vspace{0.2cm}
	\noindent\textit{Exact-Match Scoring.} This collection of methods measures the relevance of an author $a\in A^q$ with respect to the query $q$ as a function of the \textit{frequency} of $E_q$'s entities which occur in $a$'s documents. More precisely, an author $a \in A^q$ is first retrieved as candidate expert of $q$ if her \textit{WEM} profile contains {\em at least one} of the entities annotated in $E_q$; and then, she is ranked by means of one of the techniques reported in~Table~\ref{table:semquery:exactmatch} that take into account only the frequency of the entities explicitly occurring in its \textit{WEM} profile.

	\vspace{0.2cm}
	\noindent\textit{Related Match Scoring.}	This approach aims at implementing a {\em semantic scoring} of the authors in $A^q$, by evaluating the pertinence of the expertise of an author $a \in A^q$ according to the \textit{relatedness} among the entities in her \textit{WEM} profile and the entities in $E_q$ (as opposite to the frequency used by the previous scoring functions). Table~\ref{table:semquery:relatedmatch} reports the list of techniques used to design such a kind of {\em semantic scores}. They exploit either the structure of the graph $G_a$ (i.e. {\tt aer} and {\tt raer}) or compute the cosine similarity between the embedding vectors of the compared entities (i.e. {\tt aes}).
	
	\vspace{0.2cm}\noindent \textbf{Combining Document-Centric and Profile-Centric Strategies.} Document and profile-centric strategies are then eventually combined via proper data fusion techniques which are listed in Table~\ref{table:datafusionscore}. We designed those techniques as adaptations of the proposals in \cite{Fox94combinationof,Macdonald:2006:VCA:1183614.1183671} suitable for the experts finding problem.

	\begin{table}[t!]
		\centering
		\caption{Author-scoring techniques based on \textit{Related-Match} of entities and used by \wiser{} within its profile-centric strategy. The top-$k$ entities of author $a$ are the ones with the highest relevance score in $G_a$. In the experiment we have set $k = 0.1 \cdot |A_e|$, thus taking the top $10\%$ entities mentioned in $a$'s documents.}
		\label{table:semquery:relatedmatch}
		\begin{tabular}{llp{5.1cm}}
			\toprule
			\textbf{Name} & \textbf{Equation} & \textbf{Description} \\
			\hline
			\addlinespace \texttt{aer} & $  \frac{1}{{k\; |E_q|}} {\sum_{e_{q} \in E_q} {\sum_{i=1}^{k}  \rho_{e_{a, i},a} \cdot \operatorname{rel}(e_{q}, e_{a, i})   } } $ & Author Entity Relatedness score among the top-$k$ entities of $a$ and the entities $ e_q \in E_q$. \\
			\addlinespace \texttt{raer} & $  \frac{1}{{k\; |E_q|}} {\sum_{e_{q} \in E_q} {\sum_{i=1}^{k}  \rho_{e_{a, i},a} \cdot \operatorname{rel}(e_{q}, e_{a, i}) \cdot f( r_{a, e_{a, i}} )  } } $ & Ranked Author Entity Relatedness score that extends \texttt{aer}  with entities' relevance score. $f(r_{a,e})$ is a scaling function described in the experiments. \\ 
			\addlinespace \texttt{aes} & $ \operatorname{cosine}(  \sum_{e_q \in E_q} \vec{v}_{e_q} \;\cdot\; \vec{v}_{a,k}   ) $ & Author Entity Similarity that computes the cos-similarity between the  embedding $\vec{v}_{e_q}$ of entity $e_q \in E_q$ and the  embedding $v_{a,k}$ of author $a$. \\  \bottomrule
		\end{tabular}
	\end{table}

	\begin{table}[t!]
		
		\centering
		\caption{Data fusion techniques used by \wiser{} to combine $h$ scores (document-centric and profile-centric) of an author $a$ into a unique value that reflects the pertinence of $a$'s  expertise with the user query $q$.}
		\label{table:datafusionscore}
		\begin{tabular}{llp{8.5cm}}
			\toprule
			\textbf{Name} & \textbf{Equation} & \textbf{Description} \\
			\hline
			\addlinespace \texttt{combsum} & $\sum_{i=1}^h s_i(q, a)$  & The final score is the sum of the scores. \\
			
			\addlinespace  \texttt{combmin} & $ \min_{i=1}^h s_i(q, a)$ & The final score is the minimum of the scores.\\
			
			\addlinespace  \texttt{combmax}  & $ \max_{i=1}^h s_i(q, a) $ & The final score is the maximum between  the scores.\\
			
			\addlinespace \texttt{rrm} & $\prod_{i=1}^{h} \frac{1}{\operatorname{rank}_{i}(q, a)}$  & The final score is the product of the inversed ranking scores. \\
			
			\addlinespace  \texttt{rrs} & $ \frac{1} {\sum_{i=1}^{h} \operatorname{rank}_{i}(q, a)} $ & The final score is the inverse of the sums of the ranking scores.\\  \bottomrule
			
		\end{tabular}
	\end{table}

	\subsection{Optimization and Efficiency Details}
	
	\wiser{} implements three main algorithmic techniques that speed-up the retrieval of experts, thus making the query experience user-friendly.
	
	\vspace{0.2cm}
	\noindent \textbf{Double Index.} \wiser{}'s index is implemented with two different data structures, namely, two inverted lists that store both the association author-entities and entity-authors. This allows to efficiently retrieve at query time all the information that are needed for ranking authors with profile-centric strategies.
	
	\vspace{0.2cm}
	\noindent \textbf{Ordered Entities by Relevance Score.} Some profile-centric strategies, namely \texttt{aer} and \texttt{raer}, need to retrieve the top-$k$ most related entities of an author with respect to $E_q$, but this latter set of entities is known only at query time. This could be a slow process when dealing with many authors and many entities, so \wiser{} pre-computes and stores for each author the ordered list of her entities sorted by their relevance score, computed by means of a Personalized PageRank over $G_a$ (i.e. $a$'s \textit{WEM} profile). The computation of the top-$k$ entities in $E_a$ with respect to $E_q$ then boils down to a fast computation of a list intersection.
	
	\vspace{0.2cm}
	\noindent \textbf{Relatedness Cache.} The indexing phase of \wiser{} needs to compute the graph $G_a$ for every author $a$ of the input dataset. This could be a very slow process in the presence of many authors $a$ and many entities in $E_a$, because $G_a$ is a graph of up to $\Theta(|E_a|^2)$ edges which have to be weighted by querying the RESTful service underlying the \textsc{WikipediaRelatedness} framework. In order to speed up this computation, \wiser{} caches the edge weights as soon as they are computed. This way, if two entities occur in many subsequent graphs $G_a$, their computation is saved by accessing their cached values.

	\section{Validation}
	\label{sec:experimental-results}
	
	In order to evaluate the efficacy of \wiser{} we have set up a sophisticated experimental framework that has systematically tested the various document-centric and profile-centric strategies described in Tables \ref{table:synquery}--\ref{table:semquery:relatedmatch} and the data fusion techniques described in Tables \ref{table:datafusionscore} over the publicly available TU dataset~\cite{deRijke:tu-collection}. From these experiments we will derive the best combination of techniques that, then, will be used to compare the resulting \wiser{} against the state-of-the-art systems currently known to solve the expert finding problem.

	\subsection{Dataset}
	
	The TU~\cite{deRijke:tu-collection} dataset\footnote{We thank Christophe Van Gysel for providing us the dataset.} is an updated version of the UvT dataset, developed at Tilburg University (TU). It is currently the largest dataset available for benchmarking academia expert finding solutions, containing both Dutch and English documents. TU dataset comes with five different (human assessed) ground-truths, named from \textbf{GT1} to \textbf{GT5}. In our experiments we have decided to use \textbf{GT5} because it is considered the most recent and complete ground-truth (see \cite{deRijke:tu-collection} for details) and because it is the dataset used in the experiments of \cite{DBLP:journals/corr/GyselRW16}. Table \ref{table:dataset:tu:overview-1} offers a high-level overview about the dataset, while Table~\ref{table:dataset:tu:overview-2} offer a finer description.

	\begin{table}[t!]
		\centering
		\caption{Overview of the TU dataset \cite{DBLP:journals/corr/GyselRW16}.}
		\begin{tabular}{lr}
			\toprule
			\textbf{Resource} & \textbf{Count} \\ \hline
			\addlinespace Documents & 31,209 \\
			Author Candidates\footnotemark[7] & 977 \\
			Queries (GT5) & 1266 \\
			Document-candidate associations & 36,566 \\
			Documents with at least one associated candidate  & 27,834 \\
			Associations per document\footnotemark[8] & $1.13 \pm  0.39$ \\
			Associations per candidate & $37.43 \pm 61.00$ \\ \bottomrule
		\end{tabular}%
		\label{table:dataset:tu:overview-1}%

\vspace{0.5cm}

\caption{Document composition for the TU dataset.}
		\begin{tabular}{llll}
			\toprule
			\multirow{2}{*}{\textbf{Resource}} & \multirow{1}{*}{\textbf{Documents with}} & \multirow{1}{*}{\textbf{Documents with}} & \multirow{1}{*}{\textbf{Total num.}} \\
			& \multirow{1}{*}{\textbf{at least one author}}  &  \multirow{1}{*}{\textbf{no authors}} &  \multirow{1}{*}{\textbf{documents}} \\
			\hline
			\addlinespace Theses & $5152$ & $871$ & $6023$  \\
			Papers & $21120$ & $2504$ & $23624$ \\
			Profile pages (UK) & $495$   & $0$ & $495$ \\
			Profile pages (NL) & $524$  & $0$ & $524$  \\
			Course pages & $543$  & $0$ & $543$  \\
			Total documents & $27834$ & $31209$ & $3375$ \\ \bottomrule
		\end{tabular}%
		\label{table:dataset:tu:overview-2}%

\vspace{0.5cm}
		\centering
		\caption{Space occupancy of \wiser{}'s index built on the TU dataset.}
		\label{table:wiser-resource-space}
		\begin{tabular}{lll}
			\toprule
			\textbf{Resource}        & \textbf{Space} \\ \hline
			\addlinespace Raw Documents       & 25 MB \\
			\textsc{Elasitcsearch} Index & 40 MB \\
			\textit{WEM} Profiles (total)     & 94 MB \\
			\textit{WEM} Profiles (average per author) & 100 KB\\
			\bottomrule
		\end{tabular}
	\end{table}

	\vspace{0.2cm}
	
	\noindent \textbf{Indexing TU with \wiser{}.} Since TU dataset contains both Dutch and English documents, we {\em normalize} the data collection by translating Dutch documents into English via the tool \texttt{Translate Shell}\footnote{An open source command-line translator via Google Translate APIs.}. Then, the dataset is indexed with \wiser{}, as described in Section \ref{sec:wiser-index-contruction}. Table~\ref{table:wiser-resource-space} reports the memory occupancy of the final indexes.

	\subsection{Evaluation Metrics}
	
	In our experiments we will use the following ranking metrics that are available in the \texttt{trec\_eval} tool\footnote{\url{https://github.com/usnistgov/trec_eval}}, and are commonly used to evaluate expert-finding systems.

	\vspace{0.2cm}
 
	\noindent \textbf{Precision at \textit{k} (P@\textit{k}).} It is the fraction of retrieved authors that are relevant for a given query $q$ with respect to a given cut-off \textit{k} which considers only the topmost $k$ results returned by the evaluated system:

    \footnotetext{Only candidates with at least one single document association are considered.}
	\addtocounter{footnote}{1}
	\footnotetext{Only documents with at least one candidate association are considered.}
	\addtocounter{footnote}{1}
    
	\begin{equation}
	\operatorname{P@\textit{k}}(q) = \dfrac{|\{\mbox{relevant authors for $q$}\}\cap\{\mbox{ top-\textit{k}  retrieved authors for $q$}\}|}{k}
	\end{equation}
	
	%
	%

	\vspace{0.2cm}
	\noindent \textbf{Mean Average Precision (\textsc{MAP}).} Precision and recall are set-based measures, thus they are computed on unordered lists of authors. For systems that return ranked results, as the ones solving the expert-finding problem, it is desirable to consider the order in which the authors are returned. The following score computes the average of P@\textit{k} over the relevant retrieved authors.

	\begin{equation}
	\operatorname{AveP}(q) = \frac{\sum_{k=1}^n P@\textit{k}(q) \times \operatorname{rel_q}(k)}{|\{\mbox{relevant authors for $q$}\}|}\end{equation}
	
	where $n$ is the number of retrieved authors, $\operatorname{rel}_q(k)$ is function which equals to $1$ if the item at rank $k$ is a relevant author for $q$, $0$ otherwise. 
	
	\noindent The following score averages AveP over all queries in $Q$.
	
	\begin{equation}
	\operatorname{MAP} = \frac{\sum_{q\in Q} \operatorname{AveP}(q)}{|Q|}
	\end{equation}
	
	\vspace{0.2cm}
	%
	%
	
	\noindent \textbf{Mean Reciprocal Rank (\textsc{MRR}).} The reciprocal rank of a query response is the inverse of the rank of the first correct answer (i.e. relevant author for $q$), namely: 
	
	\begin{equation}
	{\operatorname{rec\_rank(q)}}={\frac{1}{pos(q)}}
	\end{equation}
	
	\noindent The following score averages the reciprocal rank over all queries in $Q$:
	
	\begin{equation}
	{\operatorname{MRR}}={\frac {1}{|Q|}}\sum _{{q \in Q}}{rec\_rank(q)}
	\end{equation}
	
	\vspace{0.2cm}
	%
	%
	
	\noindent \textbf{Normalized Discounted Cumulative Gain (\textsc{NDCG}).} Assuming to have a relevance score for each author, given a query $q$, we wish to have measures that give more value to the relevant authors that appear high in the ranked list of results returned for $q$ \cite{Jarvelin:2002:CGE:582415.582418}. Discounted Cumulative Gain is a measure that penalizes highly relevant authors appearing lower in the result list for $q$. This is obtained by reducing their relevance value (i.e. $rel_q()$, see above) by the logarithmic of their position in that list.
	
	\begin{equation}
	\operatorname{DCG}_{k}(q) = \operatorname{rel_q}(1) + \sum_{i=2}^{k} \frac{ {\operatorname{rel_q}(i)} }{\log_2{i}}
	\end{equation}
	
	The final measure we introduce for our evaluation purposes is among the most famous ones adopted for classic search engines \cite{Manning:2008:IIR:1394399}. It is computed by normalizing $DCG$ with respect to the best possible ranking for a given query $q$. More precisely, for a position $k$, the Ideal Discounted Cumulative Gain ($IDCG_k(q)$) is obtained by computing the $DCG_k(q)$ on the list of authors sorted by their relevance score wrt $q$. Then the measure $NDCG_k(q)$  is obtained as the ratio between $DCG_k(q)$ and $IDCG_k(q)$:
	
	\begin{equation}
	\operatorname{NDCG}_{k}(q)= \frac {\operatorname{DCG}_{k}(q)} { \operatorname{IDCG}_{k}(q)}
	\end{equation}

	\subsection{Experiments}
	
	Section \ref{sec:wiser} has described several possible techniques that \wiser{} can use to implement its document-centric and profile-centric strategies. In this section we experiment all these proposals by varying also their involved parameters. More precisely, for the document-centric strategies we experiment different document rankings and investigate also several data-fusion techniques that allow us to assign one single score to each candidate expert given all of its documents that are pertinent with the input query (see Tables \ref{table:synquery} and \ref{table:datafusionscore}). For the profile-centric strategies, we experiment the \textit{exact-} and \textit{related-}match scoring methods summarized in Tables \ref{table:semquery:exactmatch} and \ref{table:semquery:relatedmatch}. At the end, from all these figures we derive the best possible configurations of \wiser{}, and then compare them against the state-of-the-art approaches \cite{DBLP:journals/corr/GyselRW16}. This comparison will allow us to eventually design and implement a state-of-the-art version of \wiser{} that further improves the best known results, by means of a proper orchestration of document-centric, profile-centric and data-fusion strategies. Finally, in the last part of this Section we will conclude the experiments with a run-time evaluation and a qualitative analysis that will show how the combination of document- and profile-centric strategies does not only improve the quality of the returned results, but it also does not significantly alter the latency response of the system.

	\vspace{0.2cm}
	\noindent \textbf{Evaluation of the Document-Centric Strategies.} We configure \wiser{} to first rank documents via various scoring functions: i.e. \texttt{tf-idf}, \texttt{BM25}, or \texttt{Language Modeling} with \texttt{Dirichlet} or \texttt{Jelinek-Mercer} smoothing. Then, we compute a score for each author that combines two or more of the previous rankings via one of the data-fusion techniques described in Section~\ref{sec:wiser-query-time} and summarized in Table \ref{table:datafusionscore}. As far as  the smoothing configurations for \texttt{Dirichlet} or \texttt{Jelinek-Mercer} approaches are concerned, we set $\mu=2000$  and $\lambda=0.1$, as suggested by the documentation of \textsc{Elasticsearch}.
	
\begin{figure}[t!]
		\centering
		\includegraphics[scale=0.28]{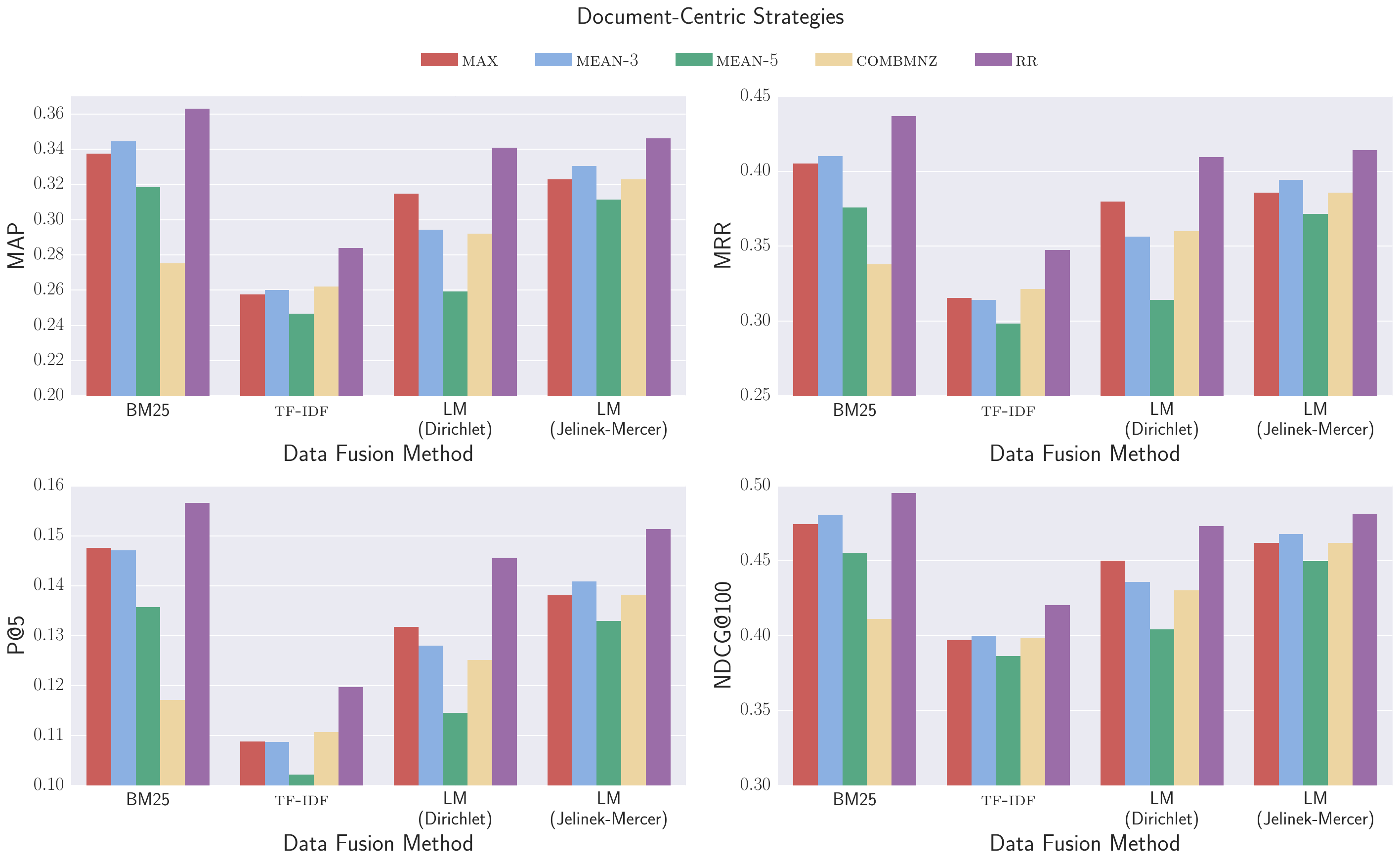}
		\caption{Expert finding performance of \wiser{} with different configurations of document-centric strategies and data-fusion methods. } 
		\label{fig:doc-cent}
		
	\end{figure}

	Figure~\ref{fig:doc-cent} reports the performance of \wiser{} by varying: (i) the document ranking, (ii) the data fusion method, and (iii) the evaluation metric. Looking at the histograms, it is very clear that each strategy achieves the best performance when the reciprocal rank (\texttt{rr} in the Figures) is used as data-fusion method. So, we have set \texttt{rr} in our following experiments and explored the best performance for all other combinations. Results are reported in Table~\ref{table:doc-cent-rr-comparison} below. We notice that, despite all strategies have values of $P@5$ and $P@10$ very close to each other, a difference is present on $MAP$, $MRR$ and $NDCG@100$. As far as the document-rankings are concerned we note that \texttt{tf-idf} is the worst approach, whereas both \texttt{LM} strategies have good performance and, undoubtly, \texttt{BM25} is the clear winner with +$7.9\%$ on $MAP$,  +$9\%$ on $MRR$ and  +$7.5\%$ on $NDCG@100$ with respect to \texttt{tf-idf}, and +$1.7\%$ on $MAP$ and +$2.3\%$ on $MRR$ and +$1.4\%$ on $NDCG@100$ with respect to any \texttt{LM}. So the winner among the document-centric strategies is {\tt BM25} with {\tt rr} as data-fusion method.

	\begin{table}[th!]
		\centering
		\caption{Comparison among different configurations of document-centric strategies with normalized reciprocal rank (\texttt{rr}) as data-fusion technique.}
		
		\label{table:doc-cent-rr-comparison}
		
		\begin{tabular}{lrrrrrrrrr}
			\toprule
			\textbf{Method} & \textbf{MAP} & \textbf{MRR} & \textbf{P@5} & \textbf{P@10} & \textbf{NDCG@100} \\
			\hline
			
			\texttt{tf-idf (rr)} & 0.284 &           0.347  &              0.120    &  0.082  &            0.420 \\
			\texttt{BM25 (rr)} & \textbf{0.363} &  \textbf{0.437}  &  \textbf{0.157} &  \textbf{0.099} &  \textbf{0.495}\\
			\texttt{LM (Dirichlet, rr)}  & 0.341 &            0.410 &            0.145   &     0.096 &           0.473  \\
			\texttt{LM (Jelinek-Mercer, rr)} & 0.346 &           0.414  &           0.151 &           0.098  &           0.481\\
			
			\bottomrule
		\end{tabular}%
		
	\end{table}%

	\newpage
    \noindent \textbf{Evaluation of the Profile-Centric Strategies.} We experimented the two configurations of \wiser{} that deploy either the \textit{exact-} or the \textit{related-}match score for evaluating the pertinence of the \textit{WEM} profile of an author with respect to the entities of a given query, as described in  Section \ref{sec:wiser-query-time}. To ease the reading of the following experimental results, we will first comment on their individual use and then illustrate some combinations.

	\begin{figure}[t!]
		\centering
		
		\includegraphics[scale=0.38]{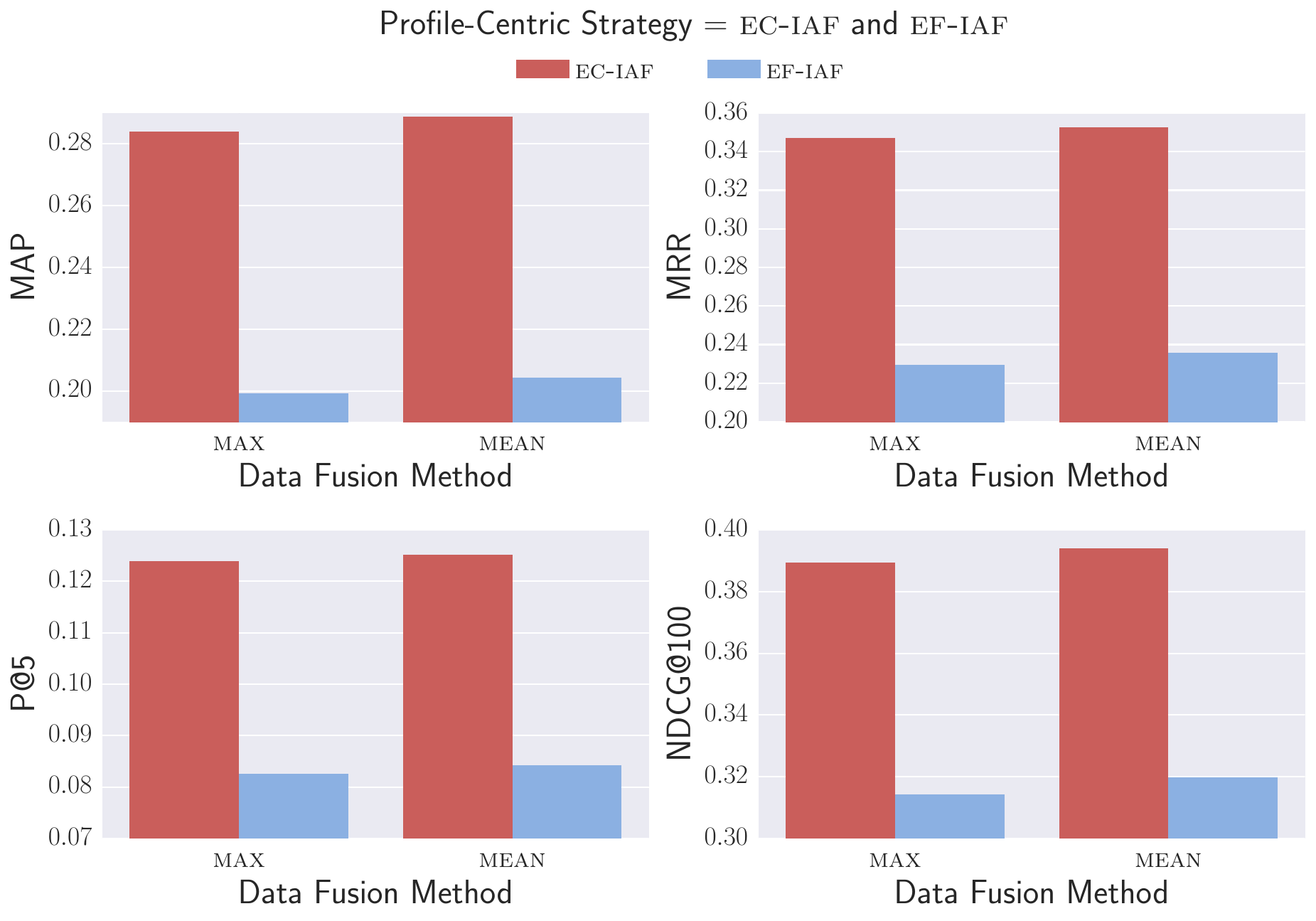}
		\vspace{-0.3cm}
		\caption{\label{fig:prof-cent-e-iaf} Performance of \wiser{} by profile-centric strategies based on entity count: \texttt{ec-iaf} and \texttt{ef-iaf}.} 

        \vspace{0.8cm}
		\includegraphics[scale=0.36]{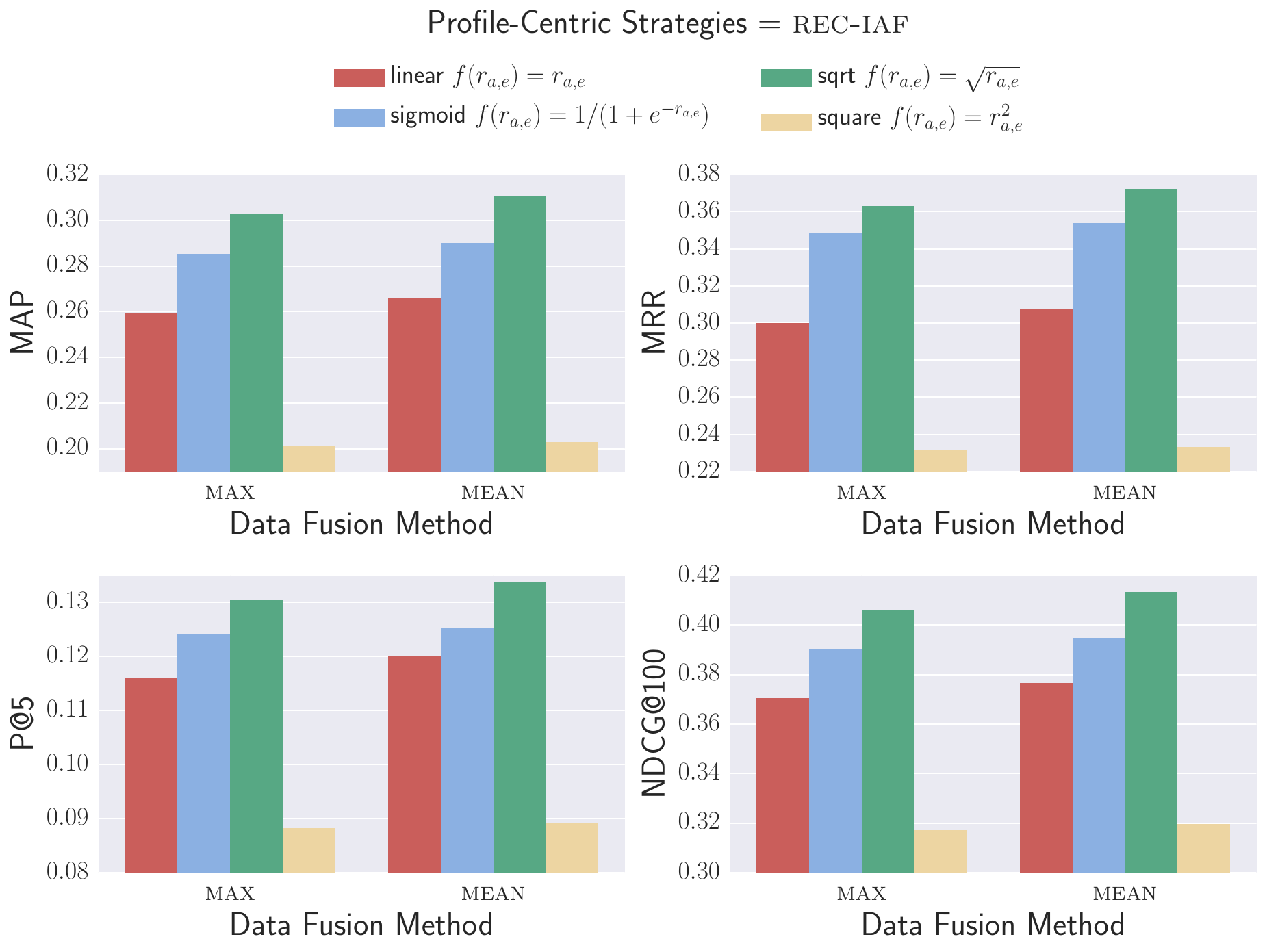}
		\vspace{-0.3cm}
		\caption{\label{fig:prof-cent-rec-iaf}. Performance of \wiser{} by \texttt{rec-iaf} as profile-centric strategy with different scaling functions $f(r_{a,e})$ for the relevance score $r_{a,e}$.} 
		
	\end{figure}

	\vspace{0.2cm}
	\noindent \textit{Exact-Match Scoring.} Figure~\ref{fig:prof-cent-e-iaf} reports the performance of \wiser{} configured to rank authors either with \texttt{ec-iaf} or \texttt{ef-iaf} (both methods based on entity frequency) and by deploying \texttt{max} and \texttt{mean} methods for combining multiple scores into a single one. It is evident that \texttt{ec-iaf} scoring with \texttt{mean} outperforms \texttt{ef-iaf}.
	
	Figure~\ref{fig:prof-cent-rec-iaf} shows the performance of \wiser{} with different configurations of \texttt{rec-iaf} scoring, which extends \texttt{ec-iaf} with the entities' relevance score $r_{a,e}$ (computed by means of Personalized PageRank executed over the author's graph $G_a$).  Since \texttt{rec-iaf} depends on $f(r_{a,e})$, we experimented various settings for $f$ that we report on the top of Figure~\ref{fig:prof-cent-rec-iaf}, i.e. identity function, sigmoid function, square root function, and square function. Looking at the plots, it is evident that the best configuration for \texttt{rec-iaf}  is achieved when $f$ is the square root function, it improves both \texttt{ec-iaf} or \texttt{ef-iaf}.

\begin{figure}[h!]
		\centering
		\includegraphics[scale=0.40]{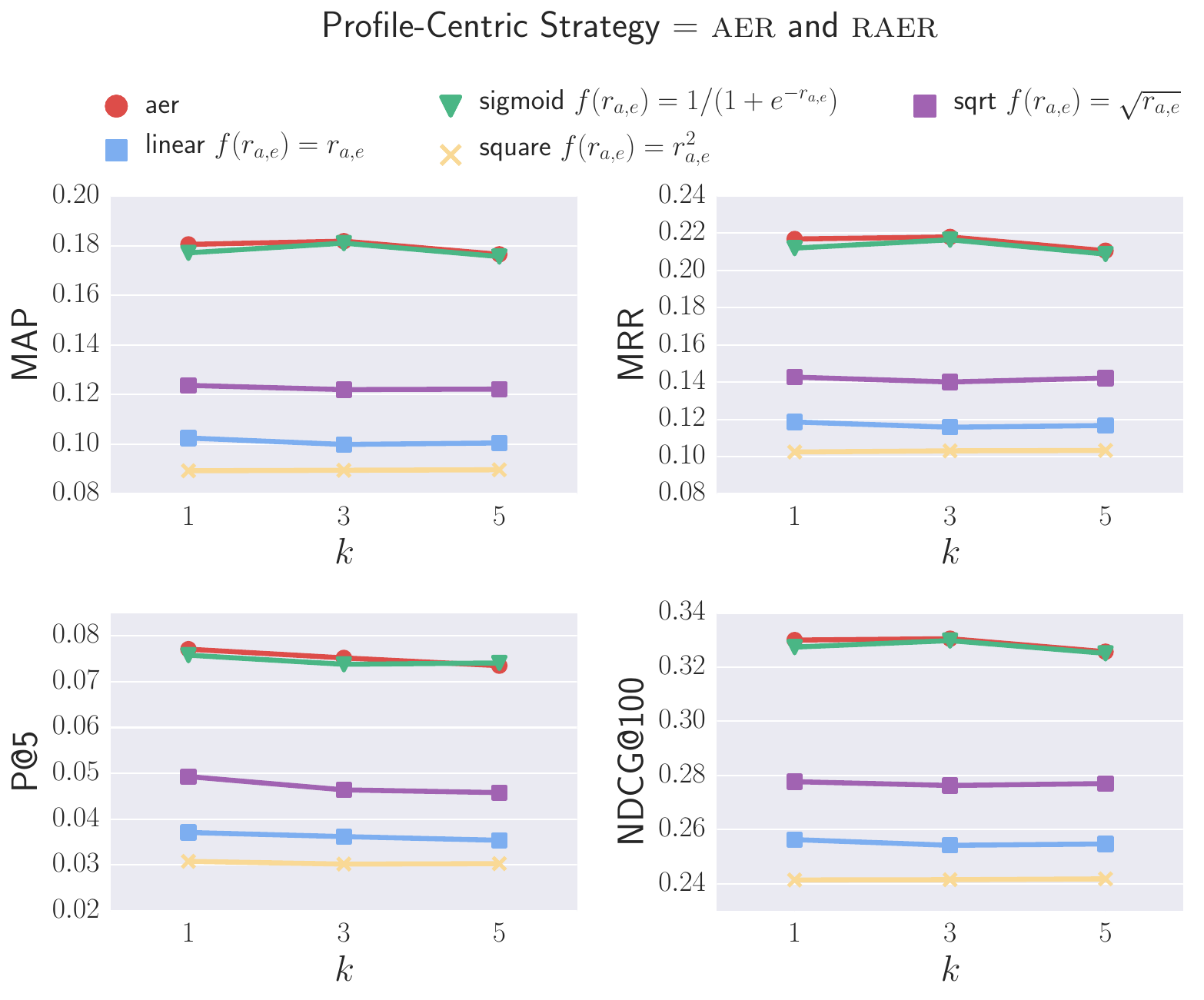}
		\vspace{-0.3cm}
		\caption{\label{fig:prof-cent-rels.eps}. Expert finding performance of \wiser{} by deploying \texttt{aer} and different configurations of \texttt{raer}.} 
	\end{figure}

	\medskip \noindent \textit{Related-Match Scoring.} Figure~\ref{fig:prof-cent-rels.eps} shows the performance of \texttt{aer} and \texttt{raer} profile-centric strategies. Since \texttt{raer} depends on $f(r_{a,e})$, we have investigated the same set of scaling functions experimented for the \texttt{rec-iaf} method. Despite the fact that the \texttt{raer} method works slightly better when configured with the sigmoid function, the simpler \texttt{aer} method is equivalent or slightly better on all metrics.
	
	\begin{figure}[h!]
		\centering
		\includegraphics[scale=0.42]{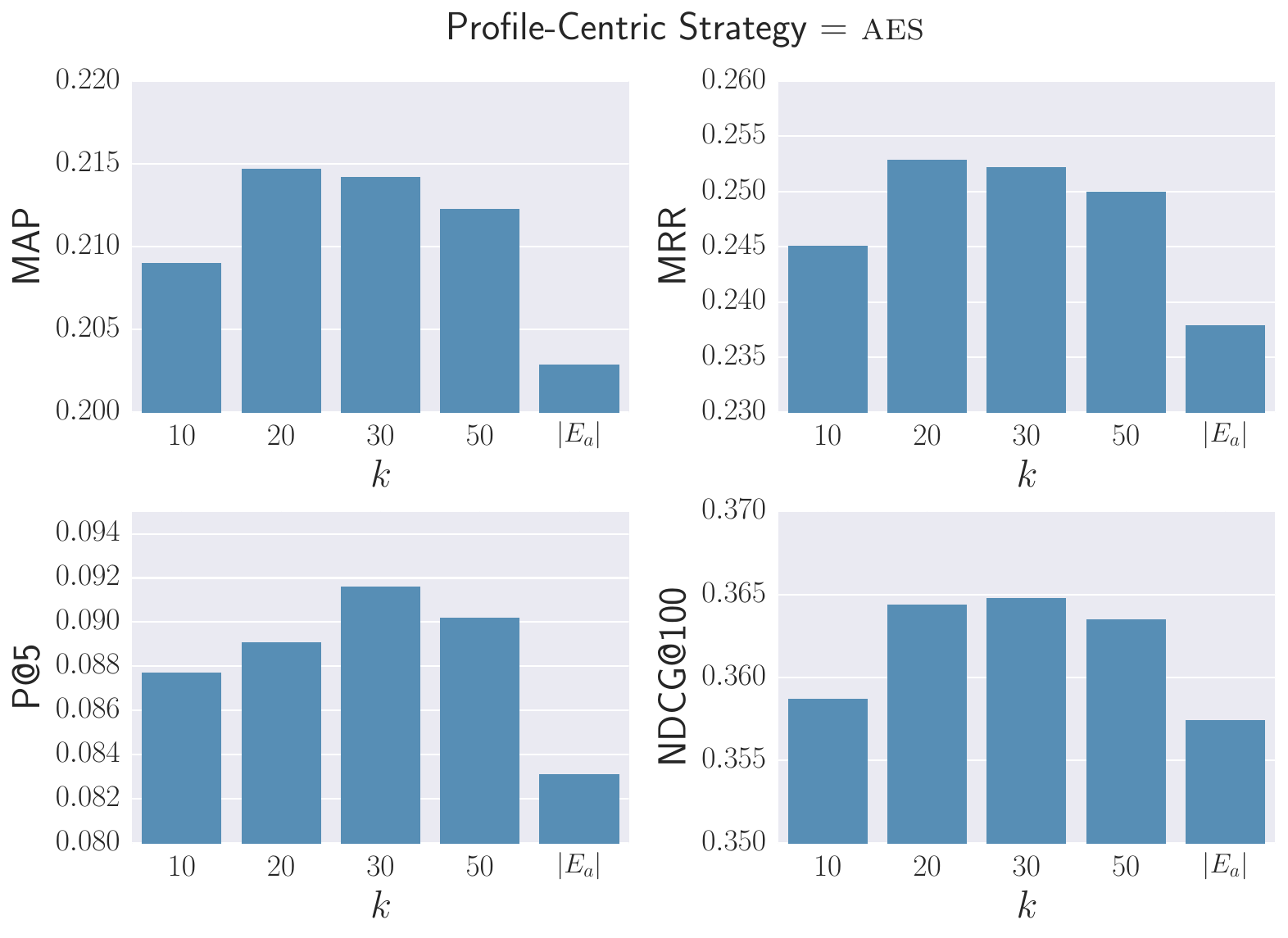}
		\vspace{-0.3cm}
		\caption{\label{fig:prof-cent-rec-aes}. Expert finding performance of \wiser{} by varying the number of top-$k$ entities used for generating the unique embedding vector of each author with \textsc{DeepWalk}-\textsc{CBOW} model.} 
	\end{figure}

	Figure~\ref{fig:prof-cent-rec-aes} reports the performance of \wiser{} which ranks authors according to \textsc{DeepWalk} embeddings models, which have been learned via \textsc{CBOW} algorithm and by fixing the size of the vectors to $100$.  In those experiments we have also evaluated the impact of varying the number $k$ of top-$k$ entities selected per author. As the plots show, ranking experts with respect to the \textsc{DeepWalk} embedding achieves better performance on different metrics and is more robust with respect to the $k$ parameter. In the following experiments we have set $k=30$. For the sake of completeness, we mention that we have also investigated the application of \textsc{DeepWalk} \textsc{Skip-gram} and \textsc{Entity2Vec}~\cite{ni2016semantic} (both \textsc{CBOW} and \textsc{Skip-gram}) models, but for the ease of explanation we did not report them since their performance are lower than \textsc{DeepWalk-CBOW}.

	\vspace{0.2cm}
	\noindent \textit{Final Discussion.} Table~\ref{table:profile-centric-best-methods} reports the best configuration found for each profile-centric method, as derived from the previous Figures. Generally speaking, methods based on \textit{exact}-match perform better than the ones based on \textit{related}-match on the TU dataset, with \texttt{rec-iaf} that achieves a peak of  +$9.7\%$ on $MAP$ with respect to the \texttt{aes} method. It is crucial to stress at this point the role of the $WEM$ profile of an author $a$ in achieving these results. In fact, the best methods --- i.e. for the \textit{exact}-match (i.e. {\tt rec-iaf}), the \textit{related}-match (i.e. {\tt aes}) and the embeddings --- are properly the ones that strongly deploy the weighted graph $G_a$ to derive, via a Personalized PageRank computation, the relevance scores $r_{a,e}$ for the entities $e$ mentioned within $a$'s documents and the corresponding top-$k$ entities.

	\begin{table}[t!]
		\centering
		
		\caption{Comparison between the best configuration of the profile-centric approaches (with both exact and related match scoring methods) implemented by \wiser{}.}
		\label{table:profile-centric-best-methods}%
		\begin{tabular}{llccccc}
			\toprule
			\textbf{Match} & \textbf{Method} & \textbf{MAP} & \textbf{MRR} & \textbf{P@5} & \textbf{P@10} & \textbf{NDCG@100} \\
			\hline\addlinespace 
			\multirow{3}{*}{\textit{Exact}} & \texttt{ec-iaf (mean)} & 0.289 &      0.353 &           0.125 &           0.081   &           0.394 \\
			
			& \texttt{ef-iaf (mean)} & 0.204 &   0.236 &       0.084  &      0.064   &            0.320 \\
			& \texttt{rec-iaf (sqrt-mean)} & \textbf{0.311} &  \textbf{0.372} &  \textbf{0.134}  &  \textbf{0.086}  &  \textbf{0.413} \\

			\hline\addlinespace 
			\multirow{3}{*}{\textit{Related}} & \texttt{aer}  & 0.187  & 0.226  & 0.081 & 0.058 & 0.332  \\
			& \texttt{raer (sigmoid)}  & 0.185 & 0.224 & 0.081 & 0.058 & 0.331 \\
			& \texttt{aes (dw-cbow-30)} & 0.214 & 0.255 & 0.092 &  0.067 & 0.365 \\
			
			\bottomrule
		\end{tabular}
	\end{table}%

	\vspace{0.4cm}
	\noindent \textbf{\wiser{} versus the State-of-the-Art.} In this last paragraph we compare the best configurations of \wiser{}, based on document- and profile-centric methods, against the best known approaches present in literature, i.e. \texttt{Log-liner}~\cite{DBLP:journals/corr/GyselRW16} and \texttt{Model 2 (jm)}~\cite{Balog:formal-models-expert-finding}.
	
	\begin{table}[h!]
		\centering
		
		\caption{Comparison between the best approaches reported in literature (top) and \wiser{}'s variants (bottom). Statistical significance of \texttt{BM25 (rr)} is computed using a two-tailed paired t-test with respect to \texttt{rec-iaf (sqrt-mean)} and indicated with $^\blacktriangle$ when $p < 0.01$. }
		\label{table:final-comparison}%
		\begin{tabular}{lllllll}
			\toprule
			\textbf{Method} & \textbf{MAP} & \textbf{MRR} & \textbf{P@5} & \textbf{P@10} & \textbf{NDCG@100} \\
			
			\hline
			\addlinespace 
			\texttt{Model 2 (jm)} \cite{Balog:formal-models-expert-finding} & 0.253 &  0.302&  0.108 & 0.081 &  0.394 \\
			\texttt{Log-linear} \cite{DBLP:journals/corr/GyselRW16} & 0.287  & 0.363 & 0.134 & 0.092 & 0.425 \\
			\hline
			\addlinespace

			\texttt{BM25 (rr)}  & \textbf{0.363}$^\blacktriangle$  &  \textbf{0.437}$^\blacktriangle$  &  \textbf{0.157}$^\blacktriangle$   &  \textbf{0.099}$^\blacktriangle$  &  \textbf{0.495}$^\blacktriangle$  \\
			\texttt{rec-iaf (sqrt-mean)}  & 0.311 &  0.372 &  0.134  & 0.086  &  0.413 \\
			\texttt{aes (dw-cbow-30)} & 0.214 & 0.255 & 0.092 &  0.067 & 0.365 \\
			
			\bottomrule
		\end{tabular}
	\end{table}

	\vspace{0.2cm}
	Table \ref{table:final-comparison} shows that both \texttt{BM25} and \texttt{rec-iaf} methods outperform \texttt{Log-linear} and \texttt{Model 2 (jm)} over different metrics. Specifically, \texttt{rec-iaf} achieves competitive performance with an improvement of +$2.4\%$ over the $MAP$ and +$0.9\%$ over $MRR$ scores with respect to \texttt{Log-linear}, whereas \texttt{BM25} improves all known methods over all metrics: +$7.6\%$ on $MAP$, +$7.4\%$ on $MRR$, +$2.3\%$ on $P@5$, +$0.7\%$ on $P@10$  and +$7\%$ on $NDCG@100$, thus resulting the clear winner and showing that for the TU dataset the document-centric strategy is better than the profile-centric strategy in \wiser{}.

	\begin{figure}[t!]
		\centering
		\includegraphics[scale=0.37]{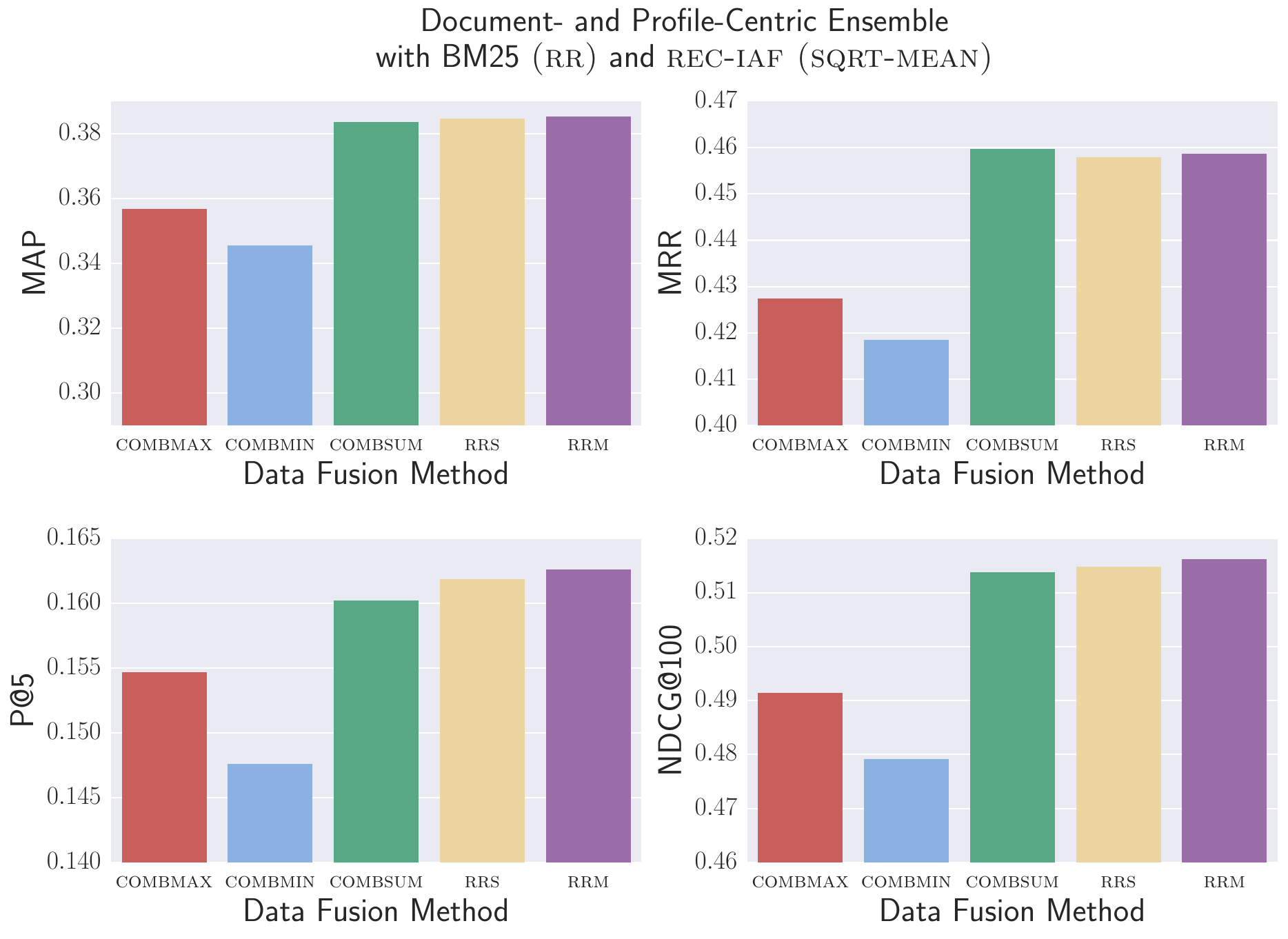}
		\vspace{-0.3cm}
		\caption{\label{fig:ensamble-bm25-rec-iaf.eps}. Performance of \wiser{} configured to combine document-centric (i.e. \texttt{BM25 (rr)}) and a profile-centric strategy (i.e. \texttt{rec-iaf (sqrt-mean)}) by means of several data-fusion techniques.} 
		
		\vspace{0.8cm}
		\includegraphics[scale=0.32]{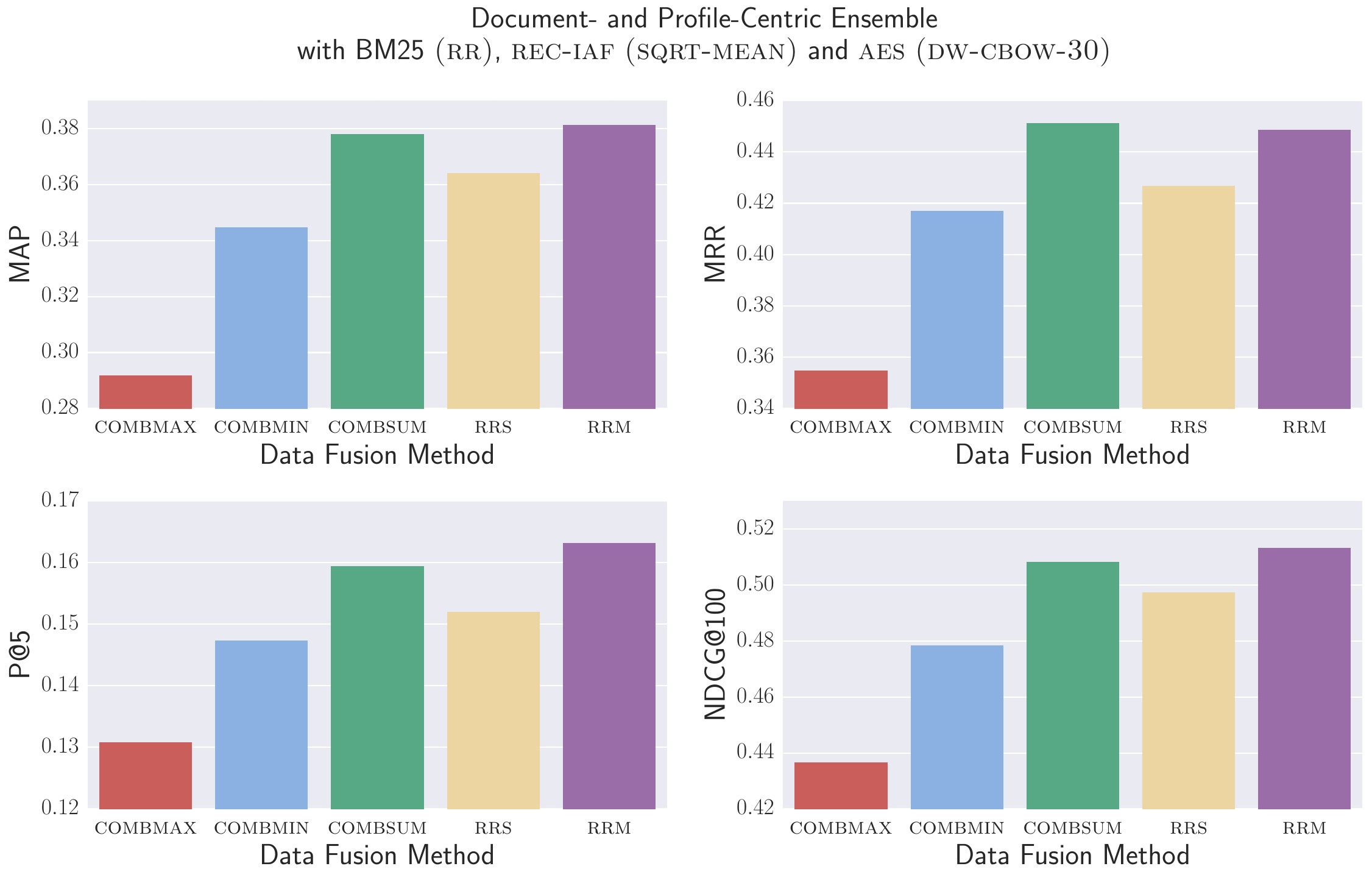}
		\vspace{-0.3cm}
		\caption{\label{fig:ensamble-bm25-rec-iaf-aes}.  Performance of \wiser{} configured to combine document-centric (i.e. \texttt{BM25 (rr)}), the best exact profile-centric strategy  (i.e. \texttt{rec-iaf (sqrt-mean)}) and the best related profile-centric strategy (i.e. \texttt{aes (dw-cbow-30)}) by means of several data-fusion techniques. } 
	\end{figure}
	
	\vspace{0.2cm}
	Given these numbers, we set up a final experiment that aimed at evaluating the best performance achievable by the combination of these methods via data-fusion techniques. Specifically, we designed a version of \wiser{} that combines the best document-centric strategy, i.e. \texttt{BM25 (rr)}, with the two best profile-centric strategies, i.e. \texttt{rec-iaf} and \texttt{aes}. Figures~\ref{fig:ensamble-bm25-rec-iaf.eps} and \ref{fig:ensamble-bm25-rec-iaf-aes} report the performance of these combinations.  The best performance are always reached when the methods at hands are combined with the \texttt{rrm} data-fusion method (purple bar).
	
	Table~\ref{table:final-comparison-ensamble} reports the performance achieved by the best known and new approaches proposed in this paper. For the sake of comparison, we also report the \texttt{Ensemble} method developed by \cite{DBLP:journals/corr/GyselRW16}, which combines via reciprocal rank (i.e. {\tt rr}) the \texttt{Log-linear} model with \texttt{Model 2 (jm)}. It is evident from the Table that 
	\begin{itemize}
		
		\item the \texttt{BM25 (rr)} implemented by \wiser{} outperforms the \texttt{Ensemble} method of~\cite{DBLP:journals/corr/GyselRW16}, which is currently the state-of-the-art, of $+3.2\%$, $+3.5\%$ and $1.8\%$ in $MAP$, $MRR$ and $NDCG@100$, and
		
		\item with a proper combination of this document-centric strategy with  the two best profile-centric algorithms of \wiser{} we are able to achieve a further improvement over \texttt{Ensemble}  on $MAP$, $MRR$ and $NDCG@100$  of  $+5.4\%$, $+5.7\%$, $+0.7\%$ and $+3.9\%$, respectively.
	\end{itemize}
	
	\noindent Therefore, \wiser{} turns out to be the new state-of-the-art solution for the expert finding problem in the academia domain.

	\begin{table}[t!]
		\centering
		\caption{Comparison between single methods (top) and different ensemble techniques whose ranking are combined via \texttt{rrm} data-fusion method. Statistical significance is computed using a one-tailed paired t-test with respect to \texttt{BM25 (rr)} (the best method of Table~\ref{table:final-comparison}) and indicated with $^\vartriangle$ for $p < 0.1$) and  $^\blacktriangle$ for $p < 0.05$).}
		\label{table:final-comparison-ensamble}%
		\resizebox{\textwidth}{!}{  
			\begin{tabular}{lllllll}
				\toprule
				\textbf{Method} & \textbf{MAP} & \textbf{MRR} & \textbf{P@5} & \textbf{P@10} & \textbf{NDCG@100} \\
				\hline
				
				\addlinespace
				\texttt{Model 2 (jm)} \cite{Balog:formal-models-expert-finding} & 0.253 &  0.302&  0.108 & 0.081 &  0.394 \\
				\texttt{Log-linear} \cite{DBLP:journals/corr/GyselRW16} & 0.287  & 0.363 & 0.134 & 0.092 & 0.425 \\
				
				\texttt{BM25 (rr)}  & {0.363} &  {0.437} & {0.157}  & {0.099} &  {0.495}  \\
				\texttt{rec-iaf (sqrt-mean)}  & 0.311 &  0.372 &  0.134  & 0.086  &  0.413  \\
				\texttt{aes (dw-cbow-30)} & 0.214 & 0.255 & 0.092 &  0.067 & 0.365 \\

				\hline

				\addlinespace 
				\texttt{Ensemble} \cite{DBLP:journals/corr/GyselRW16} & 0.331  & 0.402 & {0.156} & \textbf{0.105} & 0.477 \\

				\texttt{rrm(BM25 (rr), rec-iaf (sqrt-mean)) } & \textbf{0.385}$^\vartriangle$ & 	\textbf{0.459}$^\vartriangle$	 & \textbf{0.163} & 	0.104	 &  \textbf{0.516}$^\blacktriangle$ \\
				
				\texttt{rrm(BM25 (rr), rec-iaf (sqrt-mean), aes (dw-cbow-30)) } & 0.381$^\vartriangle$ & 	0.449	 & \textbf{0.163}	 & \textbf{0.105}$^\vartriangle$	 & 0.513$^\vartriangle$ \\

				\bottomrule
			\end{tabular}
		}
		
	\end{table}

\vspace{0.2cm}
\noindent\textbf{Run-Time Evaluation.} We supplement the large-scale quantitative evaluation described in the previous paragraph with a run-time evaluation performed on the top of the three best configurations of \wiser{}. All tests that we report here were performed on an Intel Core i7-4790 clocked at 3.60GHz, with 16GB of RAM and running Linux 4.13.

Unfortunately, we can not compare the speed of \wiser{} against to the other known systems --- i.e. \texttt{Model 2 (jm)} \cite{Balog:formal-models-expert-finding}, \texttt{Log-linear} \cite{DBLP:journals/corr/GyselRW16} and \texttt{Ensemble} \cite{DBLP:journals/corr/GyselRW16} --- because they are not publicly available. 

\begin{figure}[t]
\centering
\includegraphics[scale=0.3]{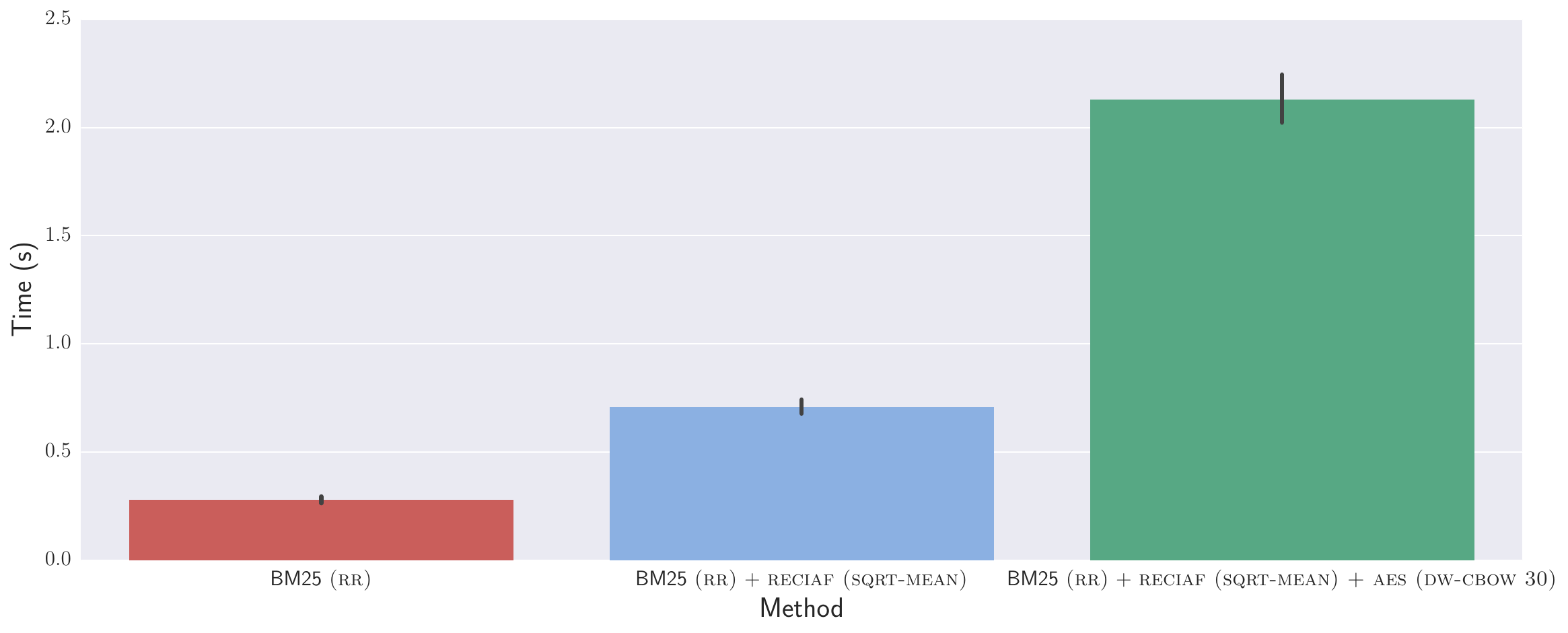}
\caption{Average run-time performance of the three best configurations of \wiser{}, executed on the $1266$ queries provided by the TU dataset. }
\label{fig:run-time}
\end{figure}

As far as the time efficiency of the three best configurations of \wiser{} is concerned --- i.e. \texttt{BM25 (rr)}, \texttt{rrm (BM25 (rr), rec-iaf (sqrt-mean))} and \texttt{rrm (BM25 (rr), rec-iaf (sqrt-mean), aes (dw-cbow-30))}, our experiments on the TU dataset show that \wiser{} is able to increase significantly its output quality but at a time cost which results not negligible for its third tested configuration. As expected, Figure~\ref{fig:run-time} shows that the simplest scoring strategy (i.e. \texttt{BM25 (rr)}) is the fastest one, since it needs only to retrieve the relevant documents for a given query and then compute the \texttt{rr} score for each candidate author. On the other hand, the second configuration \texttt{rrm (BM25 (rr), rec-iaf (sqrt-mean))} improves the quality of the results returned by \texttt{BM25 (rr)} (see Table~\ref{table:final-comparison-ensamble}) at a small time penalty, which makes the overall system still able to answer queries in less than one second. The third and last technique, i.e. \texttt{rrm (BM25 (rr), rec-iaf (sqrt-mean), aes (dw-cbow-30)}), incurs in a larger running time because it is more computationally intensive given that it needs to compute several cosine similarities between the entities annotated in the input query and the top-30 entities in the authors' profiles.

Accordingly with Table~\ref{table:final-comparison-ensamble} and with the run-time performance of the three tested methods in Figure~\ref{fig:run-time}, we decided to deploy as the final configuration of \wiser{} at \url{http://wiser.d4science.org} the one which has shown both the highest qualitative and the fastest run-time performance: namely, \texttt{rrm (BM25 (rr), rec-iaf (sqrt-mean)}).

\begin{figure}
\centering
\includegraphics[scale=0.18]{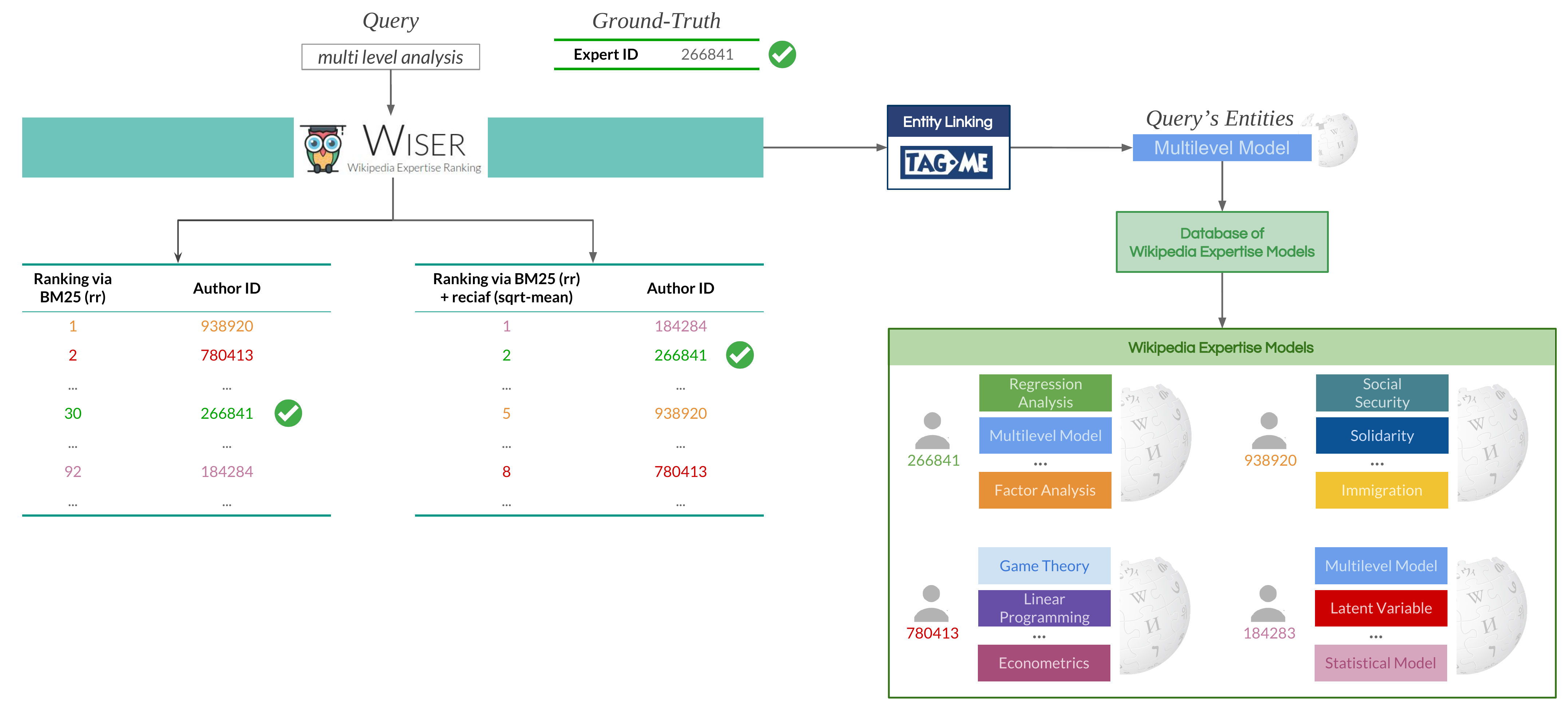}
\caption{A real example that shows where  the profile-centric approaches complement efficaciously the document-centric approaches. On the top, the input query and the ground-truth expert. On the left the final results returned by \wiser{} with two different configurations. On the right, a simplified internal architecture of the system that shows the entity annotated in the input query and a meaningful subset of entities for each ranked experts in order to show the topics concerned by their Research.}
\label{fig:example-query}
\end{figure}

\newpage
\noindent\textbf{Qualitative Analysis.}  In this last paragraph we wish to shed light on the combination between our new profile-centric approaches with the classic document-centric approaches. To fulfill this goal we have performed a qualitative analysis that consisted in manually inspecting the results returned by two configurations of \wiser{}: one instantiated with a purely document-centric approach (i.e. \texttt{BM25 (rr)}) and the other one exploiting a combination between document- and profile-centric approaches (i.e. \texttt{rrm (BM25 (rr), rec-iaf (sqrt-mean))}). We have actually identified a common pattern that recurs very frequently when the results of \texttt{BM25 (rr)} are improved by \texttt{rrm (BM25 (rr), rec-iaf (sqrt-mean))}. Figure~\ref{fig:example-query} shows one example of this common pattern for the query ``multi level analysis''. For this query \texttt{BM25 (rr)} ranks the best author (indicated in the ground truth as the one with ID equal to {\tt 266841}) at 20th position, and puts on the topmost positions many authors whose Research is unrelated to the submitted query. This worse result is due to the fact that the bag-of-words paradigm underling \texttt{BM25} incurs into the error of retrieving as experts those authors that have in their abstracts terms appearing in the query $q$ but whose meaning is different from the one intended by $q$. In fact, in the example, the words \textit{multi} and \textit{analysis} appear frequently in the papers authored by \texttt{938920} and \texttt{780413}, but with a meaning which is not related to the concept \texttt{Multilevel Model} intended by the query. Conversely,  the profile-centric score based on Wikipedia entities and adopted by \wiser{} allows to overcome this limitation. As we see in the right of Figure~\ref{fig:example-query}, \textsc{TagMe} correctly identifies the entity \texttt{Multilevel Model} both in the query $q$ and in the profile of the ground-truth expert (i.e. \texttt{266841}). On the other hand, \textsc{TagMe} does not annotate the profiles of the two non-experts, previously ranked in the top-2 positions by \texttt{BM25 (rr)} (i.e. \texttt{938920} and \texttt{780413}), with that concept. As a result, the combination of document and profile-centric approaches implemented by \texttt{rrm (BM25 (rr), rec-iaf (sqrt-mean))} allows to re-rank experts by scoring higher the ones whose profile contains the same entity of the input query $q$ (i.e. \texttt{266841}) and, vice versa, demote the ones that include the query's terms but not their corresponding concepts (i.e. \texttt{938920} and \texttt{780413}).

	\section{Conclusions}
	\label{sec:conclusions}
	
	We presented \wiser{}, a novel search engine for expert finding in academia whose novelty relies on the deployment of entity linking, plus relatedness and entity embeddings, for the the creation of the novel {\em WEM profile} for academia experts based on a weighted and labeled graph which models the {\em explicit knowledge} of author $a$ by means of the explicit (i.e. Wikipedia entities) and latent (i.e. embedding vectors) concepts occurring in her documents and their ``semantic'' relations.
	
	In the experiments we have shown that ranking authors according to the ``semantic'' relation between the user query and their \textit{WEM} profile achieves state-of-the-art performance, thus making \wiser{} the best publicly available software for academic expert finding to date.  
	
	An implementation of \wiser{} running on the Faculty of the University of Pisa is accessible at \url{http://wiser.d4science.org}. We have indexed a total of 1430 authors and 83,509 papers' abstracts, with a total of 30,984 distinct entities. Each author has published an average of 58 papers and each \textit{WEM} profile is constituted by an average of 21 unique entities. The GUI allows a user to search for a topic or for an author's name, the former returns a list of candidate experts for the queried topic, the latter returns the list of topics characterizing the \textit{WEM} profile with an estimate of their relevance. The user can browse the topics, find the papers from which they have been extracted by \wiser{}, and thus get a glimpse of the expertise of the author and, moreover, an understanding of how her research topics have evolved in the years. This tool is adopted {\em internally} to find colleagues for joint projects, and {\em externally} to allow companies and other research organizations to access in an easy way the expertise offered by our Faculty. 
	
	As a future work we foresee the exploration of: (i) other entity relatedness measures, in order to better model the edge weights in the \textit{WEM} graph; (ii) other centrality measures and clustering algorithms to estimate the relevance of an entity within an author's profile and to discard non pertinent entities; (iii) other scoring methods for the profile-centric approaches which resulted, indeed, less performing of what we expected possibly because of the noise present in the TU dataset; (iv) related to the previous point, build a new dataset for expert finding in academia or clean TU by dropping some inconsistencies we discovered in the querying process; (v) extend the use of \wiser{} to other universities and possibly explore its application to non-academia settings.

	\section*{Acknowledgments}
	\noindent We warmly thank Marco Cornolti for the preliminary processing of  the dataset and for his invaluable suggestions in the deployment of \textsc{Elasticsearch}. We also thank the anonymous reviewers for their  careful reading of the manuscript and their insightful comments that allowed us to significantly improve the quality of the paper. Part of this work has been supported by a Bloomberg Data Science Research Grant (2017), by the EU grant for the Research Infrastructure ``SoBigData: Social Mining \& Big Data Ecosystem'' (INFRAIA-1-2014-2015, agreement \#654024) and by the MIUR project FFO 2016 (DM 6 July 2016, n. 552, art. 11).

\end{document}